\documentclass[aps,superscriptaddress,dvips,11pt]{revtex4}
\usepackage{feynmp}
\usepackage{amsmath}
\usepackage{amssymb}
\usepackage{epsfig}
\usepackage{graphicx}
\usepackage{subfigure}
\usepackage{hyperref}
\usepackage{bbm}
\usepackage{color}

\newcommand{\Rmnum}[1]{\expandafter\@slowromancap\romannumeral #1@}
\allowdisplaybreaks[3]

\begin{document}

\title{A non-perturbative study of the interplay between electron-phonon interaction
and Coulomb interaction in undoped graphene}

\author{Zhao-Kun Yang}
\affiliation{Department of Modern Physics, University of Science and
Technology of China, Hefei, Anhui 230026, China}
\author{Xiao-Yin Pan}
\affiliation{Department of Physics, Ningbo University, Ningbo,
Zhejiang 315211, China}
\author{Guo-Zhu Liu}
\altaffiliation{Corresponding author: gzliu@ustc.edu.cn}
\affiliation{Department of Modern Physics, University of Science and
Technology of China, Hefei, Anhui 230026, China}

\begin{abstract}
In condensed-matter systems, electrons are subjected to two
different interactions under certain conditions. Even if both
interactions are weak, it is difficult to perform perturbative
calculations due to the complexity caused by the interplay of two
interactions. When one or two interactions are strong, ordinary
perturbation theory may become invalid. Here we consider undoped
graphene as an example and provide a non-perturbative
quantum-field-theoretic analysis of the interplay of electron-phonon
interaction and Coulomb interaction. We treat these two interactions
on an equal footing and derive the exact Dyson-Schwinger integral
equation of the full Dirac-fermion propagator. This equation depends
on several complicated correlation functions and thus is difficult
to handle. Fortunately, we find that these correlation functions
obey a number of exact identities, which allows us to prove that the
Dyson-Schwinger equation of the full fermion propagator is
self-closed. After solving this self-closed equation, we obtain the
renormalized velocity of Dirac fermions and show that its energy
(momentum) dependence is dominantly determined by the
electron-phonon (Coulomb) interaction. In particular, the
renormalized velocity exhibits a logarithmic momentum dependence and
a non-monotonic energy dependence.
\end{abstract}

\maketitle


\section{Introduction \label{Sec:introduction}}

It is sometimes necessary to study the interplay of two interactions
in condensed matter physics. For instance, disorder scattering
inevitably leads to Anderson localization \cite{Abrahams79} in
two-dimensional (2D) non-interacting metals, but direct
electron-electron interaction tends to destroy localization
\cite{Finkelstein84} and restore metallic behavior. The
metal-insulator transition found in some 2D dilute systems may
result from the interplay of disorder and electron-electron
interaction \cite{Abrahams01}. Another notable example is
phonon-mediated superconductivity. While electron-phonon interaction
(EPI) favors superconductivity by mediating an effective attraction
between electrons \cite{Schrieffer64}, direct Coulomb interaction is
repulsive and thus disfavors superconductivity. To gain a refined
description of superconductivity, one might need to consider both
EPI and Coulomb interaction.

One can employ a specific Yukawa-type fermion-boson interaction
(FBI) to describe each of the interactions mentioned above. The EPI
is already a standard FBI by definition. The Coulomb interaction can
be transformed into a Yukawa-coupling between charged electrons and
an auxiliary boson. Similar manipulation can be applied to treat
disorder scattering. In case two interactions are equally important,
one has to couple electrons to two kinds of bosons and study the
interplay of two FBIs. The coexistence of two FBIs makes theoretical
analysis rather involved. It is difficult enough to study one single
FBI, especially when its coupling constant is not small. The
traditional approach to investigate one single FBI is to adopt the
Migdal-Eliashberg (ME) theory \cite{Migdal, Eliashberg, Scalapino,
Allen, Marsiglio20}. Although ME theory was originally proposed to
treat EPI-mediated superconductivity, in the past sixty years it has
already been generalized to study many other sorts of FBIs. The
efficiency of ME theory relies crucially on the validity of Migdal
theorem \cite{Migdal}, which states that the quantum corrections to
the fermion-boson vertex function, denoted by
$\Gamma_{\mathrm{v}}(q,p)$ with $p$ ($q$) being the fermion (boson)
energy-momentum, are small and negligible. We emphasize that the
Migdal theorem is justified only in the case of weak EPI owing to
the existence of a small parameter $\lambda(\omega_D/E_F) \ll 1$,
where $\lambda$ is a dimensionless coupling constant, $\omega_D$ is
Debye frequency, and $E_F$ is Fermi energy. In a large number of
unconventional superconductors and strange metals, the reliability
of Migdal theorem and the applicability of ME theory are both in
doubt.

Recently, a non-perturbative Dyson-Schwinger (DS) equation approach
was developed by the authors \cite{Liu21, Pan21} to determine the
full fermion-boson vertex function with the help of several exact
identities. The DS equation of the full fermion propagator derived
by using this approach is self-closed and free of approximations. We
have previously applied this approach to study EPI-induced
superconducting transition in metals \cite{Liu21} and the many-body
effects caused by unscreened Coulomb interaction in Dirac fermion
systems \cite{Pan21}. Here, we generalize this approach to
investigate systems in which fermions are coupled to two different
bosons. Although our approach is generically applicable, for
concreteness we consider the interplay of EPI and Coulomb
interaction in undoped graphene \cite{Castroneto09, Sarma11,
Kotov12}. We focus on the fermion velocity renormalization induced
by such an interplay.

The impact of Coulomb interaction on the properties of graphene has
been extensively studied by meas of both perturbative expansion
method \cite{Gonzalez94, DasSarma07, Polini07, Son07, Vafek07,
Son08, Aleiner08, Foster08, Kotov08, Kotov09, Vozmediano10,
Vozmediano11, Fogler12, Mishchenko07, Vafek08, Hofmann14, Barnes14,
Throckmorton15, Sharma16} and non-perturbative method
\cite{Khveshchenko01, Gorbar02, Khveshchenko04, Khveshchenko09,
Liu09, Gamayun10, WangLiu12, WangLiu14, Gonzalez12, Gonzalez12jhep,
Popovici13, Carrington16, Carrington18}. An interesting problem is
to determine how the fermion velocity is renormalized by the Coulomb
interaction. In 1994, Gonzalez \emph{et al.} \cite{Gonzalez94}
carried out a first-order renormalization group (RG) analysis of the
Coulomb interaction by using the weak-coupling perturbation theory
and revealed a logarithmic renormalization of the fermion velocity,
described by $v \propto \ln \left(\Lambda/|\mathbf{p}|\right)$,
where $\Lambda$ is an ultraviolet cutoff of fermion momentum
$\mathbf{p}$. Experiments have observed a logarithmic velocity
renormalization \cite{Elias11, Lanzara11, Chae12}, which appears to
be qualitatively consistent with first-order RG result. Barnes
\emph{et al.} \cite{Barnes14} calculated some higher-order (two-loop
and three-loop) corrections and concluded that the logarithmic
behavior obtained at first-order is qualitatively altered by such
corrections, which signals the breakdown of weak-coupling
perturbation theory. In a recent paper \cite{Pan21}, we revisited
this problem by employing our DS equation approach and found that
the Dirac fermion velocity does exhibit a logarithmic momentum
dependence if all the interaction-induced corrections are taken into
account in a non-perturbative way.

In actual graphene materials, there are other types of interactions
than the Coulomb interaction. For instance, phonons are always
present as the consequence of lattice vibrations. Their interaction
with Dirac fermions could affect the spectral properties
\cite{Louie07, Tse07} and the transport properties \cite{Sarma11} of
graphene, and might lead to some ordering instabilities in certain
circumstances \cite{Meng19, Scalettar19}. In principle, the
renormalized velocity $v(\mathbf{p})$ observed in experiments should
receive contributions not only from Coulomb interaction but also
from EPI. It is therefore important to consider both of these two
interactions so as to make a more direct comparison between
theoretical calculations and experimental results. A particularly
interesting question is: would EPI change the logarithmic momentum
dependence of renormalized velocity caused by the Coulomb
interaction?

In this paper, we describe the interplay of EPI and Coulomb
interaction by coupling fermions to two different bosons. We first
write down an effective model for such an interplay and then derive
the DS equation of the full Dirac fermion propagator $G(p)$ within
the functional-integral formalism of quantum field theory. This DS
equation has a much more complicated expression than that generated
by one single FBI since it contains four two-point correlation
functions and two vertex functions. After making a careful analysis,
we find that these six correlation functions obey two exact
identities, which then leads to a great simplification of the DS
equation of $G(p)$. But there is still a unknown current vertex
function $\Gamma_{0}(q,p)$, where $q$ is boson momentum, in the
simplified equation. We further obtain four generalized
Ward-Takahashi identities (WTIs) and show that $\Gamma_{0}(q,p)$ can
be expressed as a linear combination of $G^{-1}(p)$ by solving these
four WTIs. Based on all of these results, we prove that the exact DS
equation of $G(p)$ is self-closed.

We then apply our approach to compute the renormalized velocity of
Dirac fermions. After numerically solving the self-closed DS
equation of $G(p)$, we obtain the energy- and momentum-dependence of
the renormalized velocity $v(\epsilon,\mathbf{p})$. Our finding is
that the energy dependence and momentum dependence of
$v(\epsilon,\mathbf{p})$ are dominantly determined by EPI and
Coulomb interaction, respectively. More concretely, EPI leads to an
obvious non-monotonic energy dependence of $v(\epsilon)$ at a fixed
$|\mathbf{p}|$. For any given $\epsilon$, $v(\mathbf{p})$ exhibits a
logarithmic $|\mathbf{p}|$-dependence over a wide range of
small-$|\mathbf{p}|$ region. A clear indication of this result is
that the logarithmic velocity renormalization caused by the Coulomb
interaction is not changed by the additional EPI.

The rest of the paper is organized as follows. In
Sec.~\ref{Sec:DSE}, we first define the effective model of the
system and then derive the DS equation of the full fermion
propagator $G(p)$ after taking into account the contributions from
two different FBIs. In Sec.~\ref{Sec:WTIs}, we derive four exact
generalized WTIs satisfied by $G(p)$ and $\Gamma_{0}(q,p)$ together
with three other current vertex functions. We show that these
identities can be used to make the DS equation of $G(p)$
self-closed. In Sec.~\ref{Sec:numerics}, we provide the numerical
solutions for $G(p)$ and analyze the influence of the interplay of
EPI and Coulomb interaction on the renormalization of fermion
velocity. We briefly summarize the main results of the paper and
discuss further research projects in Sec.~\ref{Sec:summary}. A
detailed functional analysis of the interaction vertex function and
the derivation of the DS equations of fermion and boson propagators
are presented in Appendix \ref{Sec:appa} and Appendix
\ref{Sec:appb}, respectively.

\section{Dyson-Schwinger equation of fermion propagator \label{Sec:DSE}}

The unusual physical properties of two-dimensional massless Dirac
fermions has already been widely investigated in the context of
undoped graphene \cite{Castroneto09, Sarma11, Kotov12}. The Dirac
fermions in graphene have eight indices, including two sublattices,
two inequivalent valleys, and two spin directions. To describe these
fermions, one can define a standard four-component spinor $\psi =
(c_{AK},c_{BK},c_{BK'},c_{AK'})^{T}$, where $A,B$ are sublattices
and $K,K'$ are inequivalent valleys. For such a representation, the
fermion flavor is $N=2$, corresponding to two spin components. The
dynamics of Dirac fermions can be described by the following
Lagrangian density
\begin{eqnarray}
\mathcal{\mathcal{L}} = \mathcal{L}_{f}+\mathcal{L}_{p} +
\mathcal{L}_{A} + \mathcal{L}_{fp} + \mathcal{L}_{fA},
\label{eq:originallagrangian}
\end{eqnarray}
in which the five terms are formally written as
\begin{eqnarray}
\mathcal{L}_{f} &=& \sum_{\sigma}^N \bar{\psi}_{\sigma}(x)
\left(i\partial_{0}\gamma_{0} - i\partial_{1}\gamma_{1} -
i\partial_{2}\gamma_{2}\right)\psi_{\sigma}(x), \\
\mathcal{L}_{p} &=& \frac{1}{2}\phi^\dag(x)\mathbb{D}(x)\phi(x),\\
\mathcal{L}_{A} &=& \frac{1}{2}A(x)\mathbb{F}(x)A(x),\\
\mathcal{L}_{fp} &=& - \sum_{\sigma}^{N}
g\phi(x)\bar{\psi}_{\sigma}(x)\gamma_0\psi_{\sigma}(x),\\
\mathcal{L}_{fA} &=&- \sum_{\sigma}^{N}
A(x)\bar{\psi}_{\sigma}(x)\gamma_0\psi_{\sigma}(x).
\end{eqnarray}
Here, the $4 \times 4$ matrices $\gamma_{\mu}$, where $\mu=0,1,2$,
satisfy standard Clifford algebra. $\bar{\psi}$ is defined via
$\gamma_{0}$ as $\bar{\psi} = \psi^{\dag}\gamma_{0}$. $x$ is a
three-dimensional vector, i.e., $x\equiv (x_{0},\mathbf{x}) =
(x_{0},x_{1},x_{2})$. Time $x_{0}$ can be either real or imaginary
(Matsubara time), and all the results obtained in this paper are
equally valid in both cases. Throughout this and the next sections,
we utilize a real time, i.e., $x_{0}=t$, for notational simplicity.
The subscript $\sigma$ sums from $N=1$ to $N=2$. The bare fermion
velocity $v_{F}$ is already absorbed into the spatial derivatives,
namely $v_{F}\partial_{1,2}\rightarrow
\partial_{1,2}$, which makes notations simpler. The scalar field
$\phi$ represents the phonon. $\mathcal{L}_{f}$ and
$\mathcal{L}_{p}$ are the kinetic terms of Dirac fermions and
phonons, respectively, and $\mathcal{L}_{fp}$ describes the EPI.
Originally, the Coulomb interaction between Dirac fermions is
modeled by the Hamiltonian term
\begin{eqnarray}
H_{C} = \frac{1}{4\pi}\frac{e^{2}}{v_{F}\epsilon}
\sum_{\sigma,\sigma'}^{N} \int d^2\mathbf{x} d^2 \mathbf{x}'{\bar
\psi}_{\sigma} (\mathbf{x}) \gamma_{0} \psi_{\sigma}(\mathbf{x})
\frac{1}{\left|\mathbf{x} - \mathbf{x}'\right|}{\bar
\psi}_{\sigma'}(\mathbf{x}')\gamma_{0}
\psi_{\sigma'}(\mathbf{x}'),\nonumber
\end{eqnarray}
where $e$ is the electric charge and $\varepsilon$ is the dielectric
constant whose value depends on the substrate of undoped graphene
\cite{Castroneto09, Sarma11, Kotov12}. Here we couple an auxiliary
scalar field $A$ to the spinor field $\psi$ and use $\mathcal{L}_{A}
+ \mathcal{L}_{fA}$ to equivalently describe the Coulomb interaction
\cite{Son07, Barnes14, Pan21}. Two operators $\mathbb{D}$ and
$\mathbb{F}$ are introduced to define the equations of the free
motions of $\phi$ and $A$: $\mathbb{D}\phi = 0$ and $\mathbb{F}A =
0$. Notice that the FBI terms $\mathcal{L}_{fp}$ and
$\mathcal{L}_{fA}$ do not mix different flavors since both $\phi$
and $A$ couple to the fermion density operator $\rho(x)
=\sum_{\sigma}^{N} \bar{\psi}_{\sigma}(x)\gamma_0\psi_{\sigma}(x)$.

The quantum many-body effects of graphene induced by the long-range
Coulomb interaction, which is often described by the coupling
between $\psi$ and $A$, has previously been studied by using various
field-theoretic methods. Such methods can be roughly classified into
two categories: perturbative expansion \cite{Gonzalez94, DasSarma07,
Polini07, Son07, Son08, Aleiner08, Kotov08, Kotov09, Vozmediano10,
Vozmediano11, Fogler12, Mishchenko07, Vafek08, Hofmann14, Barnes14,
Sharma16} and non-perturbative DS equation \cite{Khveshchenko01,
Gorbar02, Khveshchenko04, Khveshchenko09, Liu09, Gamayun10,
WangLiu12, WangLiu14, Gonzalez12, Gonzalez12jhep, Popovici13,
Carrington16, Carrington18}. Two parameters are frequently used to
perform perturbative series expansion, namely the fine-structure
constant $\alpha$ and the inverse of fermion flavor $1/N$. However,
as demonstrated in Ref.~\cite{Pan21}, both of these two parameters
are actually not small enough to guarantee the validity of the
perturbative expansion method. On the other hand, previous
non-perturbative DS equation calculations focused on the excitonic
pairing instability \cite{Khveshchenko01, Gorbar02, Khveshchenko04,
Khveshchenko09, Liu09, Gamayun10, WangLiu12, WangLiu14, Gonzalez12,
Gonzalez12jhep, Popovici13, Carrington16, Carrington18}. Little
effort has been devoted to computing the renormalized fermion
velocity by using the DS equation approach. It turns out that the
results obtained by different groups of authors are inconsistent
with each other (see Ref.~\cite{Pan21} for a recent review). This
inconsistency originates from the fact that the vertex corrections
have not been incorporated in a satisfactory manner in previous DS
equation studies \cite{WangLiu12, Gonzalez12, Gonzalez12jhep,
Popovici13, Carrington16, Carrington18}. In Ref.~\cite{Pan21}, we
have developed an efficient method to incorporate all the vertex
corrections to the $\psi$-$A$ coupling and adopted this method to
determine the full energy-momentum dependence of renormalized
fermion velocity without introducing any approximation.

The correlation effects induced by EPI has also been investigated in
the context of graphene-like systems \cite{Louie07, Tse07, Roy14}.
The interplay between EPI and Coulomb interaction was considered by
means of perturbative RG method \cite{Aleiner08}. To the best of our
knowledge, the non-perturbative effects of the interplay between EPI
and Coulomb interaction have not been studied previously. In this
work, we generalize the DS equation approach reported in
Ref.~\cite{Pan21} to treat the coupling of Dirac fermions to two
distinct bosons.

In order to generate various correlation functions, we now introduce
three external sources and change the original Lagrangian density
$\mathcal{L}$ to
\begin{eqnarray}
\mathcal{L}_{T} = \mathcal{L} + J\phi + K A +\sum_{\sigma}^N
\left(\bar{\psi}_{\sigma}\eta_{\sigma} +
\bar{\eta}_{\sigma}\psi_{\sigma}\right), \label{eq:totallagrangian}
\end{eqnarray}
where $J$, $K$, $\eta$, and $\bar{\eta}$ are external sources for
$\phi$, $A$, $\psi^{\dag}$, and $\psi$, respectively. The partition
function (generating functional) is
\begin{eqnarray}
Z[J,K,\bar{\eta},\eta] \equiv \int D\phi DA D\bar{\psi}_{\sigma}
D\psi_{\sigma} e^{i\int dx \mathcal{L}_{T}},
\end{eqnarray}
where $\int dx \equiv \int d^{3}x = \int dt d^{2}\mathbf{x}$. The
generating functional for connected correlation functions is defined
via $Z$ as
\begin{eqnarray}
W\equiv W[J,K,\bar{\eta},\eta] = -i\ln Z[J,K,\bar{\eta},\eta].
\end{eqnarray}
The full propagators of Dirac fermion $\psi$, phonon $\phi$, and
boson $A$ are defined in order as follows
\begin{eqnarray}
G_{\sigma}(x-y) = -i\langle\psi_{\sigma}(x)
\bar{\psi}_{\sigma}(y)\rangle = \frac{\delta^2W}{\delta
\bar{\eta}_{\sigma}(x)\delta \eta_{\sigma}(y)}\Big|_{J=0},\\
D(x-y) = -i\langle\phi(x)\phi^\dag(y)\rangle =
-\frac{\delta^2W}{\delta J(x)\delta J(y)}\Big|_{J=0},\\
F(x-y) = -i\langle A(x)A(y)\rangle = -\frac{\delta^2W}{\delta K(x)
\delta K(y)}\Big|_{J=0}.
\end{eqnarray}
Hereafter we use an abbreviated notation $J=0$ to indicate that all
external sources are taken to vanish. The propagator $G_{\sigma}$ of
each flavor has the same form, so the subscript $\sigma$ can be
omitted. There are two additional correlation functions that convert
$\phi$ and $A$ into each, defined by
\begin{eqnarray}
D_{F}(x-y)=-i\langle\phi(x)A(y)\rangle = -\frac{\delta^2W}{\delta
J(x)\delta K(y)}\Big|_{J=0}, \\
F_{D}(x-y)=-i\langle A(x)\phi(y)\rangle = -\frac{\delta^2W}{\delta
K(x) \delta J(y)}\Big|_{J=0}.
\end{eqnarray}
It is clear that $D_{F}(x-y)$ and $F_{D}(x-y)$ both vanish at the
tree-level as the model does not contain such a term as
$\phi(x)A(x)$. However, they become finite once quantum (i.e.,
loop-level) corrections are taken into account. It will become clear
that $D_{F}(x-y)$ and $F_{D}(x-y)$ make nonzero contributions to the
fermion self-energy.

For each FBI, there exists a specific interaction vertex function,
which plays an important role since it enters into the DS equation
of both fermion and boson propagators. Two FBIs naturally correspond
to two interaction vertex functions. Such vertex functions can be
generated by such correlation functions as $\langle
\phi(x)\psi(y)\bar{\psi}(z)\rangle$ and $\langle A(x)
\psi(y)\bar{\psi}(z)\rangle$. To illustrate how to define
interaction vertex functions, let us use $W$ to generate the
following connected three-point correlation function:
\begin{eqnarray}
\langle\phi(x)\psi(y)\bar{\psi}(z)\rangle_c &=&
\frac{\delta^3W}{\delta J(x)\delta\bar{\eta}(y)\delta\eta(z)}
\Big|_{J=0}.\label{eq:5}
\end{eqnarray}
Here, a subscript $c$ is introduced to indicate that the correlation
function is connected. As shown in Appendix \ref{Sec:appa}, this
correlation function can be expressed in terms of the fermion and
boson propagators as
\begin{eqnarray}
\frac{\delta^3W}{\delta J(x)\delta\bar{\eta}(y)
\delta\eta(z)}\Big|_{J=0} &=& -\int
dx^{\prime}dy^{\prime}dz^{\prime}D(x-x^{\prime})
G(y-y^{\prime})\frac{\delta^3\Xi}{\delta\phi(x^{\prime})
\delta\bar{\psi}(y^{\prime})\delta
\psi(z^{\prime})}\Big|_{J=0}G(z^{\prime}-z) \nonumber \\
&& -\int dx^{\prime} dy^{\prime} dz^{\prime} D_F(x-x^{\prime})
G(y-y^{\prime}) \frac{\delta^3\Xi}{\delta
A(x^{\prime})\delta\bar{\psi}(y^{\prime})
\delta\psi(z^{\prime})}\Big|_{J=0}
G(z^{\prime}-z),\label{eq:deltawdg} \nonumber \\
\end{eqnarray}
where the generating functional for proper (irreducible) vertices
$\Xi$ is defined via $W$ as
\begin{eqnarray}
\Xi = W - \int dx \Big[J\langle \phi\rangle + K\langle A\rangle +
\sum_{\sigma}^N\left(\bar{\eta}_{\sigma}\langle\psi_{\sigma}\rangle +
\langle\bar{\psi}_{\sigma}\rangle\eta_{\sigma}\right)\Big].
\label{eq:Xi}
\end{eqnarray}
The interaction vertex function for EPI is defined as
$$\Gamma_{p}(y-x,x-z) = \frac{\delta^3\Xi}{\delta\phi(x)
\delta\bar{\psi}(y)\delta\psi(z)}\Big|_{J=0},$$ and
that for $\psi$-$A$ coupling is defined as $$\Gamma_{A}(y-x,x-z) =
\frac{\delta^3\Xi}{\delta A(x) \delta\bar{\psi}(y)
\delta\psi(z)}\Big|_{J=0}.$$ It is necessary to emphasize
that $\Gamma_{p}$ and $\Gamma_{A}$ depend on two (not three) free
variables, namely $y-x$ and $x-z$. The propagators and interaction
vertex functions appearing in Eq.~(\ref{eq:deltawdg}) are Fourier
transformed as follows:
\begin{eqnarray}
G(p) &=& \int dx e^{ip\cdot x}G(x), \\
D(q) &=& \int dx e^{iq\cdot x}D(x), \\
D_F(q) &=& \int dx e^{iq\cdot x}D_F(x), \\
\Gamma_p(q,p) &=& \int dx dy e^{i(p+q)\cdot(y-x)}e^{ip\cdot(x-z)}
\Gamma_{p}(y-x,x-z), \\
\Gamma_A(q,p) &=& \int dx dy e^{i(p+q)\cdot(y-x)}e^{ip\cdot(x-z)}
\Gamma_{A}(y-x,x-z).
\end{eqnarray}
Here, the three-momentum is $p \equiv (p_{0},\mathbf{p}) =
(p_{0},p_{1},p_{2})$. Performing Fourier transformation to
$\langle\phi(x)\psi(y)\bar{\psi}(z)\rangle_c$, we
find
\begin{eqnarray}
&&\int dxdy e^{i(p+q)\cdot(y-x)}e^{ip\cdot(x-z)}
\langle\phi(x)\psi(y)\bar{\psi}(z)\rangle_c
\nonumber \\
&=& -D(q)G(p+q)\Gamma_p(q,p)G(p)-D_F(q)G(p+q)\Gamma_A(q,p)G(p).
\label{eq:expanded}
\end{eqnarray}

Then we replace the boson field $\phi$ with the boson field $A$ and
consider another three-point correlation function $\langle
A(x)\psi(y)\bar{\psi}(z)\rangle_c$. After carrying
out similar calculations, we obtain
\begin{eqnarray}
&&\int dx dy e^{i(p+q)\cdot(y-x)}e^{ip\cdot(x-z)}
\langle A(x)\psi(y)\bar{\psi}(z)\rangle_c\nonumber\\
&=&-F_D(q)G(p+q)\Gamma_p(q,p)G(p)-F(q)G(p+q)\Gamma_A(q,p)G(p),
\label{eq:expanded2}
\end{eqnarray}
where $F(q)$ and $F_D(q)$ are transformed from $F(x)$ and $F_D(x)$
respectively as
\begin{eqnarray}
F(q) &=& \int dx e^{iq\cdot x}F(x), \\
F_{D}(q) &=& \int dx e^{iq\cdot x}F_{D}(x).
\end{eqnarray}

In the framework of quantum field theory \cite{Itzykson}, all the
$n$-point correlation functions are connected to each other by an
infinite number of DS integral equations. The single particle
properties of Dirac fermions are embodied in the full fermion
propagator $G(p)$, which satisfies the following DS equation
\begin{eqnarray}
G^{-1}(p) &=& G_{0}^{-1}(p) - i\int dq g\gamma_0
G(p+q)D(q)\Gamma_p(q,p) - i\int dq \gamma_0 G(p+q)F(q)\Gamma_A(q,p)
\nonumber \\
&& -i \int dq g\gamma_0 G(p+q)D_F(q)\Gamma_A(q,p) - i\int dq
\gamma_0G(p+q)F_D(q)\Gamma_p(q,p). \label{eq:DSEGP}
\end{eqnarray}
Here, we introduce the abbreviation $\int dq \equiv
\frac{d^{3}q}{(2\pi)^{3}}$. The derivational details that lead to
this equation are shown in Appendix \ref{Sec:appb}.

\begin{figure}[htbp]
\centering
\includegraphics[width=3.9in]{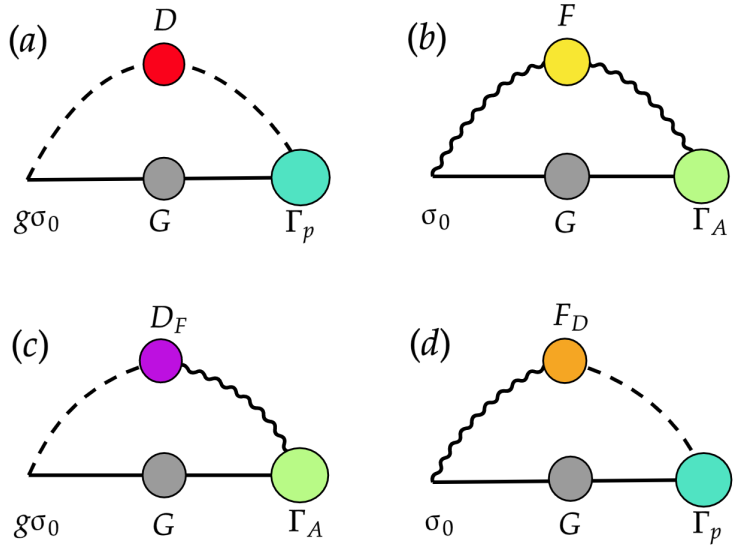}
\caption{The four diagrams (a)-(d) correspond to the four terms of
the fermion self-energy given by Eq.~(\ref{eq:DSEGP}). Dashed (wavy)
line represents the propagation of the boson $\phi$ ($A$).}
\label{fig:selfenergy4}
\end{figure}

According to Eq.~(\ref{eq:DSEGP}), the fermion self-energy
$\Sigma(p)=G^{-1}(p)-G_{0}^{-1}(p)$ consists of four terms. The
corresponding diagrams are shown in Fig.~\ref{fig:selfenergy4}. The
first two terms originate from pure EPI and pure Coulomb
interaction, respectively. The last two terms represent the
contributions from the mixing of two bosons. In previous theoretical
works, the last two terms are often naively neglected. In
Eq.~(\ref{eq:DSEGP}), there are four two-point correlation
functions, namely $D(q)$, $F(q)$, $D_{F}(q)$, and $F_{D}(q)$, and
two interaction vertex functions, including $\Gamma_{p}(q,p)$ and
$\Gamma_{A}(q,p)$. These six functions are all unknown and each of
them satisfies its own DS integral equation. According to the
analysis presented in Refs.~\cite{Liu21, Pan21}, the DS equations of
$\Gamma_{p}(q,p)$ and $\Gamma_{A}(q,p)$ are extremely complicated
since they are coupled to an infinite number of DS equations obeyed
by all the higher-point correlation functions.

\begin{figure}[htbp]
\centering
\includegraphics[width=4.6in]{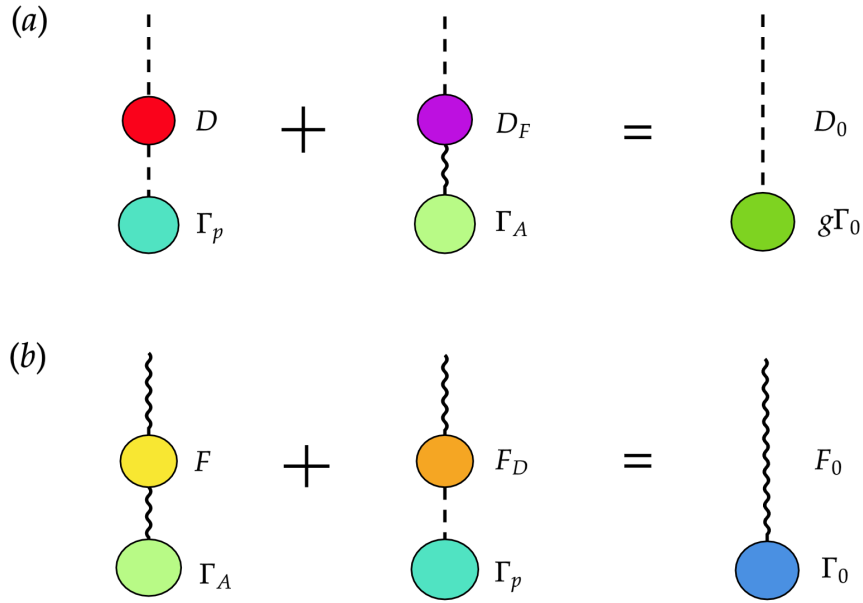}
\caption{The Feynman diagrams plotted in (a) and (b) correspond to
Eq.~(\ref{eq:d0gamma0}) and Eq.~(\ref{eq:f0gamma0}), respectively.
The free propagators $D_{0}$ and $F_{0}$ are represented by dashes
and wavy lines without carrying a shadowed circle, respectively.}
\label{fig:gammas}
\end{figure}

At first glance, the above DS equation of $G(p)$ is not self-closed
and cannot be solved because it contains six unknown functions
$D(q)$, $F(q)$, $D_{F}(q)$, $F_{D}(q)$, $\Gamma_p(q,p)$, and
$\Gamma_{A}(q,p)$. Fortunately, we find that it is not necessary to
determine each of these six functions separately. Indeed, these six
functions satisfy two exact identities. The derivation of the exact
identities is based on the invariance of partition function $Z$
under an arbitrary infinitesimal change of the scalar field $\phi$.
Such an invariance gives rise to
\begin{eqnarray}
\langle\mathbb{D}(x)\phi(x)-g\sum_{\sigma}^N
\bar{\psi}_{\sigma}(x)\gamma_0\psi_{\sigma}(x) + J\rangle = 0,
\end{eqnarray}
which is simply the mean value of the equation of the motion of
phonons. Since $\langle \phi(x) \rangle = \frac{\delta W}{J(x)}$, we
re-write this equation as
\begin{eqnarray}
\mathbb{D}(x)\frac{\delta W}{J(x)} = g\sum_{\sigma}^{N}
\langle\bar{\psi}_{\sigma}(x) \gamma_{0}\psi_{\sigma}(x)\rangle + J.
\end{eqnarray}
Then we carry out functional derivatives with respect to sources
$\eta(z)$ and $\bar{\eta}(y)$ in order. After taking all sources to
zero, we have
\begin{eqnarray}
\mathbb{D}(x) \langle\phi(x)\psi(y)
\bar{\psi}(z)\rangle_c = g\langle\sum_{\sigma}^N
\bar{\psi}_{\sigma}(x)\gamma_0 \psi_{\sigma}(x) \psi(y)
\bar{\psi}(z)\rangle_c, \label{eq:DJetaeta}
\end{eqnarray}
where Eq.~(\ref{eq:5}) is used in the calculation. In order to find
out the consequence of this equation, we need to perform a Fourier
transformation for both sides. With the help of
Eq.~(\ref{eq:expanded}), it is easy to find that the left-hand side
(l.h.s.) of Eq.~(\ref{eq:DJetaeta}) becomes
\begin{eqnarray}
D_0^{-1}(q)\Big[-D(q)G(p+q)\Gamma_p(q,p)G(p)
-D_F(q)G(p+q)\Gamma_A(q,p)G(p)\Big] \label{eq:D0DGGammaG}
\end{eqnarray}
after making Fourier transformation. The free phonon propagator
$D_{0}(q)$ is obtained by Fourier transformation of the operator
$\mathbb{D}(x)$. Then we turn to deal with the right hand side
(r.h.s.) of Eq.~(\ref{eq:DJetaeta}). It can be verified that the
Lagrangian density $\mathcal{L}$ given by
Eq.~(\ref{eq:originallagrangian}) respects a U(1) symmetry $\psi
\rightarrow e^{i\theta}\psi$, where $\theta$ is an infinitesimal
constant. Noether theorem dictates that this symmetry leads to a
conserved current $j_{\mu}\equiv \left(j_{0},j_{1},j_{2}\right)$,
satisfying the identity
\begin{eqnarray}
\sum_{\mu}\partial^{\mu}j_{\mu}(x) \equiv \partial_{0}j_{0}(x) -
\partial_{1}j_{1}(x) - \partial_{2}j_{2}(x)=0.
\label{eq:currentconservation}
\end{eqnarray}
The three components of local current operator $j_{\mu}(x)$ can be
expressed in terms of spinor field as
\begin{eqnarray}
j_{0}(x) &=& \sum_{\sigma}^N \bar{\psi}_{\sigma}(x)\gamma_{0}
\psi_{\sigma}(x), \label{eq:j0x}\\
j_{1}(x) &=& \sum_{\sigma}^N \bar{\psi}_{\sigma}(x)\gamma_{1}
\psi_{\sigma}(x), \label{eq:j1x}\\
j_{2}(x) &=& \sum_{\sigma}^N \bar{\psi}_{\sigma}(x)\gamma_{2}
\psi_{\sigma}(x).\label{eq:j2x}
\end{eqnarray}
Now the r.h.s. of Eq.~(\ref{eq:DJetaeta}) is equivalent to $g\langle
j_{0}(x)\psi(y)\bar{\psi}(z)\rangle$. Here it is
convenient to introduce a special current vertex function
$\Gamma_{0}(x-z,z-y)$ and define it via the relation
\begin{eqnarray}
\langle\sum_{\sigma}^N \bar{\psi}_{\sigma}(x)\gamma_0
\psi_{\sigma}(x)\psi(y)\bar{\psi}(z)\rangle_c = -\int d\xi
d\xi' G(y-\xi)\Gamma_{0}(\xi-x,x-\xi')G(\xi'-z).
\label{eq:currentgp0}
\end{eqnarray}
The Fourier transformation of $\Gamma_{0}(\xi-x,x-\xi')$ is given by
\begin{eqnarray}
\Gamma_{0}(\xi-x,x-\xi') = \int dq dp
e^{-i(p+q)\cdot(\xi-x)-ip\cdot(x-\xi')}\Gamma_{0}(q,p).
\label{eq:Fouriercurrentvertex}
\end{eqnarray}
Fourier transforming the r.h.s. of Eq.~(\ref{eq:DJetaeta}) leads to
\begin{eqnarray}
\int dx dy e^{i(p+q)\cdot(y-x)} e^{ip\cdot(x-z)}
g\langle\sum_{\sigma}^N \bar{\psi}_{\sigma}(x)\gamma_0
\psi_{\sigma}(x)\psi(y)\bar{\psi}(z)\rangle_c \rightarrow
-g G(p+q)\Gamma_{0}(q,p)G(p).\label{eq:gGGammaG}
\end{eqnarray}
The two formulae shown in Eq.~(\ref{eq:D0DGGammaG}) and
Eq.~(\ref{eq:gGGammaG}) must be equal, i.e.,
\begin{eqnarray}
D_0^{-1}(q)\Big[D(q)G(p+q)\Gamma_p(q,p)G(p)
+D_F(q)G(p+q)\Gamma_A(q,p)G(p)\Big] = g G(p+q)\Gamma_{0}(q,p)G(p),
\end{eqnarray}
which can be simplified to a more compact form
\begin{eqnarray}
D(q)\Gamma_p(q,p)+D_F(q)\Gamma_A(q,p) = D_0(q)g\Gamma_0(q,p).
\label{eq:d0gamma0}
\end{eqnarray}
The above analysis can be easily applied to treat the coupling
between $\psi$ and $A$. Repeating the same calculational steps gives
rise to another important identity
\begin{eqnarray}
F_D(q)\Gamma_p(q,p)+F(q)\Gamma_A(q,p) = F_0(q)\Gamma_0(q,p),
\label{eq:f0gamma0}
\end{eqnarray}
where $F_0(q)$ is the free propagator of $A$ boson, obtained by
performing Fourier transformation to the operator $\mathbb{F}(x)$.
In Fig.~\ref{fig:gammas}, we show a diagrammatic illustration of the
two identities given by Eq.~(\ref{eq:d0gamma0}) and
Eq.~(\ref{eq:f0gamma0}).

\begin{figure}[htbp]
\centering
\includegraphics[width=4.8in]{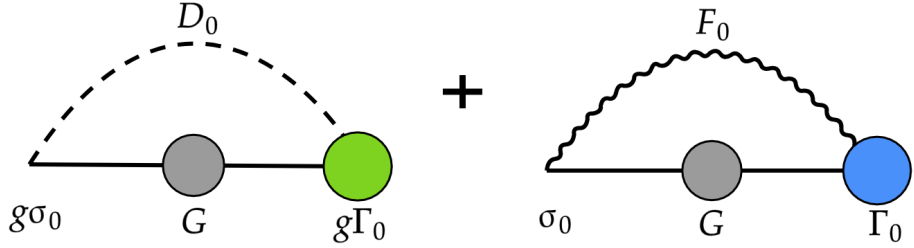}
\caption{Diagrams for the fermion self-energy appearing in the
simplified DS equation (\ref{eq:dsefinal}).}\label{fig:selfenergy2}
\end{figure}

Making use of the two identities of Eq.~(\ref{eq:d0gamma0}) and
Eq.~(\ref{eq:f0gamma0}), the originally complicated DS equation
(\ref{eq:DSEGP}) can be greatly simplified to
\begin{eqnarray}
G^{-1}(p) = G_{0}^{-1}(p) - i\int dq \left[g^2D_0(q) + F_0(q)\right]
\gamma_0 G(p+q)\Gamma_{0}(q,p).\label{eq:dsefinal}
\end{eqnarray}
The sum of the four self-energy diagrams shown in
Fig.~\ref{fig:selfenergy4} are now replaced with the sum of the two
diagrams shown in Fig.~\ref{fig:selfenergy2}. This equation looks
much simpler, but is still hard to solve since the function
$\Gamma_{0}(q,p)$ remains unknown. The equation of $G(p)$ could be
entirely self-closed if and only if $\Gamma_{0}(q,p)$ depends solely
on $G(p)$. Our next task is to find out the relationship between
$\Gamma_{0}(q,p)$ and $G(p)$.

\section{Generalized Ward-Takahashi identities \label{Sec:WTIs}}

In this section we will show that the function $\Gamma_{0}(q,p)$ can
be expressed purely in terms of $G(p)$. The calculational procedure
that leads to the exact relation between $\Gamma_{0}(q,p)$ and
$G(p)$ has previously been illustrated with great details in
Refs.~\cite{Liu21, Pan21}. Here, in order to make this paper
self-contained, we briefly outline the main calculational steps.

Now make the following global transformation to the spinor field
$\psi(x)$:
\begin{eqnarray}
\psi(x) \to e^{i\theta \gamma_{m}}\psi(x), \quad
\bar{\psi}(x)\to\psi^{\dag}(x) e^{-i\theta
\gamma_{m}}\gamma_{0}.\label{eq:u1global}
\end{eqnarray}
Here, $\theta$ is an infinitesimal constant and $\gamma_{m}$ denotes
a generic $4\times 4$ matrix. Generically, there are totally $32$
different choices for $\gamma_{m}$. $16$ of them are $\gamma_{m}=I$,
$\gamma_{m}=\gamma_{0}$, $\gamma_{m}=\gamma_{1}$,
$\gamma_{m}=\gamma_{2}$, $\gamma_{m}=\gamma_{3}$,
$\gamma_{m}=\gamma_{0}\gamma_{1}\equiv\gamma_{01}$,
$\gamma_{m}=\gamma_{0}\gamma_{2}\equiv\gamma_{02}$,
$\gamma_{m}=\gamma_{0}\gamma_{3}\equiv\gamma_{03}$,
$\gamma_{m}=\gamma_{1}\gamma_{2}\equiv\gamma_{12}$,
$\gamma_{m}=\gamma_{1}\gamma_{3}\equiv\gamma_{13}$,
$\gamma_{m}=\gamma_{2}\gamma_{3}\equiv\gamma_{23}$,
$\gamma_{m}=\gamma_{0}\gamma_{1}\gamma_{2}\equiv\gamma_{012}$,
$\gamma_{m}=\gamma_{0}\gamma_{1}\gamma_{3}\equiv\gamma_{013}$,
$\gamma_{m}=\gamma_{0}\gamma_{2}\gamma_{3}\equiv\gamma_{023}$,
$\gamma_{m}=\gamma_{1}\gamma_{2}\gamma_{3}\equiv\gamma_{123}$, and
$\gamma_{m}=\gamma_{0}\gamma_{1}\gamma_{2}\gamma_{3}\equiv\gamma_{0123}$.
The rest $16$ matrices are obtained by multiplying each of these
matrices by $i$. It should be emphasized that we do not require the
total Lagrangian density $\mathcal{L}_{T}$ defined by
Eq.~(\ref{eq:totallagrangian}) to be invariant under the above
global transformation. In fact, $\mathcal{L}_{T}$ is invariant under
the transformation (\ref{eq:u1global}) only when $\gamma_{m}=I$.
Different from $\mathcal{L}_{T}$, the partition function
$Z[J,K,\bar{\eta},\eta]$ should be invariant under the
transformation $\psi \to e^{i\theta \gamma_{m}}\psi$ for any choice
of $\gamma_{m}$, since $Z[J,K,\bar{\eta},\eta]$ is obtained by
integrating out all the possible configurations of $\psi$ and
$\psi^{\dag}$.

Below we will demonstrate that the invariance of
$Z[J,K,\bar{\eta},\eta]$ under the infinitesimal transformation
Eq.~(\ref{eq:u1global}) imposes a stringent constraint on the
relation between $\Gamma_{0}(q,p)$ and $G(p)$. Making use of this
invariance, we derive the following equation
\begin{eqnarray}
&&\langle\big(i\partial_0\bar{\psi}_{\sigma}(x) \gamma_0 -
i\partial_1 \bar{\psi}_{\sigma}(x)\gamma_1 -
i\partial_2\bar{\psi}_{\sigma}(x)\gamma_2\big)\gamma_{m}
\psi_{\sigma}(x)\rangle + \langle \psi_{\sigma}^\dag(x)
\gamma_{m}^{\dag}\gamma_0 \big(i\partial_0\gamma_0 -
i\partial_{1}\gamma_{1} - i\partial_{2}\gamma_{2}\big)
\psi_{\sigma}(x)\rangle \nonumber \\
&=& \bar{\eta}_{\sigma}(x)\gamma_{m} \langle \psi_{\sigma}(x)\rangle
-\langle\psi_{\sigma}^{\dag}(x)\rangle
\gamma_{m}^{\dag}\gamma_0 \eta_{\sigma}(x) \nonumber \\
&& -g\langle\phi(x)\psi_{\sigma}^{\dag}(x)
(\gamma_0\gamma_0\gamma_{m}-\gamma_{m}^{\dag}\gamma_0\gamma_0)
\psi_{\sigma}(x)\rangle - \langle A(x)\psi_{\sigma}^{\dag}(x)
(\gamma_0\gamma_0\gamma_{m}-\gamma_{m}^{\dag}\gamma_0\gamma_0)
\psi_{\sigma}(x)\rangle,\label{eq:variationaction}
\end{eqnarray}
which comes from the identity $\delta Z = 0$. Throughout this
section, the repeated flavor index $\sigma$ needs to be summed over.
But we omit the summation notation for simplicity. As the next step,
we carry out functional derivatives $\frac{\delta^2}{\delta
\bar{\eta}(y)\delta\eta(z)}|_0$ to both sides of
Eq.~(\ref{eq:variationaction}) and obtain
\begin{eqnarray}
&&\langle \big(i\partial_0\bar{\psi}_{\sigma}(x)\gamma_{0} -
i\partial_{1}\bar{\psi}_{\sigma}(x)\gamma_{1} -
i\partial_{2}\bar{\psi}_{\sigma}(x)\gamma_{2}\big)
\gamma_{m}\psi_{\sigma}(x)\psi(y)\bar{\psi}(z)\rangle_{c} \nonumber \\
&& +\langle\psi_{\sigma}^{\dag}(x)\gamma_{m}^{\dag}
\gamma_0\big(i\partial_{0}\gamma_{0} -
i\partial_{1}\gamma_{1}-i\partial_{2}\gamma_{2}\big)
\psi_{\sigma}(x)\psi(y)\bar{\psi}(z) \rangle_{c} \nonumber \\
&=& \delta(x-y)\gamma_{m}G(x-z) - G(y-x)\gamma_0\gamma_{m}^{\dag}
\gamma_0 \delta(x-z) \nonumber \\
&& - g\langle \phi(x)\psi_{\sigma}^\dag(x)(\gamma_0\gamma_0
\gamma_{m} - \gamma_{m}^{\dag}\gamma_0\gamma_0)\psi_{\sigma}(x)
\psi(y)\bar{\psi}(z)\rangle_{c}
\nonumber \\
&& -\langle A(x)\psi_{\sigma}^\dag(x)(\gamma_0\gamma_0\gamma_{m} -
\gamma_{m}^{\dag}\gamma_0\gamma_0)\psi_{\sigma}(x)\psi(y)
\bar{\psi}(z)\rangle_{c}.\label{eq:fivepoint}
\end{eqnarray}
While this formula is strictly valid, it is formally too
complicated. In particular, the third term of the r.h.s. is a very
special correlation function defined by the mean value of the
product of five field operators. The forth term has a similar
structure. The presence of such special correlation functions makes
it difficult to extract useful information on the relation between
$\Gamma_{0}(q,p)$ and $G(p)$. Fortunately, it is easy to see that
these two five-point correlation functions can be eliminated if the
matrix $\gamma_{m}$ is properly chosen to ensure that
$\gamma_0\gamma_0\gamma_{m} - \gamma_{m}^{\dag}\gamma_0\gamma_0 =
0$. Let us choose the following four matrices
\begin{eqnarray}
\gamma_{m}=I, \quad \gamma_{m}=\gamma_{01}, \quad
\gamma_{m}=\gamma_{02}, \quad \gamma_{m}=i\gamma_{12}.
\end{eqnarray}
Substituting them into Eq.~(\ref{eq:fivepoint}) eliminates the third
and the forth terms of the r.h.s. of this equation, leaving us with
an identity of the form
\begin{eqnarray}
&&\langle \big(i\partial_0\bar{\psi}_{\sigma}(x)\gamma_{0} -
i\partial_{1}\bar{\psi}_{\sigma}(x)\gamma_{1} -
i\partial_{2}\bar{\psi}_{\sigma}(x)\gamma_{2}\big)
\gamma_{m}\psi_{\sigma}(x)\psi(y)\bar{\psi}(z)\rangle_{c} \nonumber \\
&& +\langle\psi_{\sigma}^{\dag}(x)\gamma_{m}^{\dag}
\gamma_0\big(i\partial_{0}\gamma_{0} -
i\partial_{1}\gamma_{1}-i\partial_{2}\gamma_{2}\big)
\psi_{\sigma}(x)\psi(y)\bar{\psi}(z) \rangle_{c} \nonumber \\
&=& \delta(x-y)\gamma_{m}G(x-z) - G(y-x)\gamma_0\gamma_{m}^{\dag}
\gamma_0 \delta(x-z).\label{eq:currentgp}
\end{eqnarray}

For $\gamma_{m} = I$, the identity of Eq.~(\ref{eq:currentgp})
becomes
\begin{eqnarray}
&&\langle\big(i\partial_0\bar{\psi}_{\sigma}(x) \gamma_{0} -
i\partial_{1}\bar{\psi}_{\sigma}(x)\gamma_{1} -
i\partial_2\bar{\psi}_{\sigma}(x)\gamma_{2}\big)
\psi_{\sigma}(x)\psi(y)\bar{\psi}(z)\rangle_{c} \nonumber \\
&& +\langle \bar{\psi}_{\sigma}(x)\big(i\partial_{0}\gamma_{0} -
i\partial_{1}\gamma_{1}-i\partial_{2}\gamma_{2}\big)
\psi_{\sigma}(x)\psi(y)\bar{\psi}(z) \rangle_{c} \nonumber \\
&=& \delta(x-y)IG(x-z) - G(y-x)I\delta(x-z).\label{eq:currentgpI}
\end{eqnarray}
Using the conserved current operator $j_{\mu}(x)$ define by
Eqs.~(\ref{eq:j0x}-\ref{eq:j2x}), we find that
Eq.~(\ref{eq:currentgpI}) can be re-written as
\begin{eqnarray}
i\partial^{\mu}\langle j_{\mu}(x)\psi(y) \bar{\psi}(z)\rangle_{c}
&\equiv& i\partial_{0}\langle j_{0}(x)\psi(y) \bar{\psi}(z)\rangle_{c} -
i\partial_{1}\langle j_{1}(x)\psi(y) \bar{\psi}(z)\rangle_{c} -
i\partial_{2}\langle j_{2}(x)\psi(y) \bar{\psi}(z)\rangle_{c} \nonumber
\\
&=& \delta(x-y)G(x-z)-G(x-y)\delta(x-z).
\label{eq:divergencecurrent}
\end{eqnarray}
The three correlation functions appearing in the l.h.s. of this
equation are used to define three current vertex functions
$\Gamma_{0,1,2}$ as follows
\begin{eqnarray}
\langle j_{0,1,2}(x)\psi(y)\bar{\psi}(z)\rangle_{c} = -\int d\xi
d\xi'G(y-\xi)\Gamma_{0,1,2}(\xi-x,x-\xi')G(\xi'-z).
\label{eq:currentvertex1}
\end{eqnarray}
The function $\Gamma_{0}$ has already been encountered in in
Sec.~\ref{Sec:DSE}, and its Fourier transformation is given by
Eq.~(\ref{eq:Fouriercurrentvertex}). The other two functions
$\Gamma_{1}$ and $\Gamma_{2}$ can be transformed similarly, namely
\begin{eqnarray}
\Gamma_{1,2}(\xi-x,x-\xi') = \int dq dp
e^{-i(p+q)\cdot(\xi-x)-ip\cdot(x-\xi')}\Gamma_{1,2}(q,p).
\label{eq:Fouriercurrentvertex12}
\end{eqnarray}
The next step would be to substitute Eq.~(\ref{eq:currentvertex1})
into Eq.~(\ref{eq:divergencecurrent}) and carry out Fourier
transformation to both sides of Eq.~(\ref{eq:divergencecurrent}).
The calculation is straightforward. For instance,
$i\partial_0\langle j_{1}(x)\psi(y)\bar{\psi}(z)\rangle_{c}$ can be
Fourier transformed as follows
\begin{eqnarray}
&&i\partial_0\langle j_{1}(x)\psi(y)\bar{\psi}(z)\rangle_{c}
\nonumber \\
&=& -i\partial_0 \int d\xi d\xi'
G(y-\xi)\Gamma_{0}(\xi-x,x-\xi')G(\xi'-z) \nonumber \\
&=&-i\partial_0\int d\xi d\xi'\int dp dq dp' dq'
e^{-i(p+q)\cdot(y-\xi)}G(p+q)
e^{-i(p'+q')\cdot(\xi-x)-ip'\cdot(x-\xi')} \Gamma_{0}(q',p')
e^{-ip\cdot(\xi'-z)}G(p) \nonumber \\
&=&-i\partial_0\int dp dq dp' dq' e^{-i(p+q)\cdot y}G(p+q)
\delta\big(p+q-(p'+q')\big)e^{i(p'+q')\cdot x - ip'\cdot x}
\Gamma_{0}(q',p')
\delta(p'-p)e^{ip\cdot z}G(p)\nonumber \\
&=&-i\partial_0\int dp dq e^{-i(p+q)\cdot(y-x)} e^{-ip\cdot(x-z)}
G(p+q)\Gamma_{0}(q,p) G(p) \nonumber \\
&=&\int dp dq e^{-i(p+q)\cdot(y-x)}e^{-ip\cdot (x-z)} q_{0}
G(p+q)\Gamma_{0}(q,p) G(p).\label{eq:Fouriersplitting1}
\end{eqnarray}
After completing all the analytical calculations, we eventually
convert Eq.~(\ref{eq:divergencecurrent}) into
\begin{eqnarray}
G(p+q)\big[q_{0}\Gamma_{0}(q,p)-q_{1}\Gamma_{1}(q,p) -
q_{2}\Gamma_{2}(q,p)\big] G(p) = G(p)-G(p+q),
\end{eqnarray}
which can be further simplified to
\begin{eqnarray}
q_{0}\Gamma_{0}(q,p)-q_{1}\Gamma_{1}(q,p)-q_{2}\Gamma_{2}(q,p) =
G^{-1}(p+q)-G^{-1}(p). \label{eq:WTI1}
\end{eqnarray}
Recall this identity is derived by making the transformation $\psi
\to e^{i\theta}\psi$ and $\bar{\psi}\to e^{-i\theta}\bar{\psi}$,
which is nothing but the global U(1) symmetry
of the Lagrangian density. Thus this identity is indeed the ordinary
WTI induced by the conservation of particle number.

As demonstrated at the end of Sec.~\ref{Sec:DSE}, the DS equation of
the full fermion propagator $G(p)$, given by
Eq.~(\ref{eq:dsefinal}), would be made entirely self-closed if we
could express the function $\Gamma_{0}(q,p)$ purely in terms of
$G(p)$. Apparently, it is not possible to entirely determine
$\Gamma_{0}(q,p)$ by solving the above WTI, since $\Gamma_{1}(q,p)$
and $\Gamma_{2}(q,p)$ are also unknown. To determine
$\Gamma_{0}(q,p)$, we need to find our more identities satisfied by
$\Gamma_{0}(q,p)$, $\Gamma_{1}(q,p)$, $\Gamma_{2}(q,p)$, and $G(p)$.

Next we choose $\gamma_{m} = \gamma_{01}$ and use this matrix to
express the identity of Eq.~(\ref{eq:currentgp}) in the form
\begin{eqnarray}
&&\langle\big(i\partial_0\bar{\psi}_{\sigma}(x) \gamma_{1} -
i\partial_{1}\bar{\psi}_{\sigma}(x)\gamma_{0} -
i\partial_2\bar{\psi}_{\sigma}(x)\gamma_{0}\gamma_{1}\gamma_{2}\big)
\psi_{\sigma}(x)\psi(y)\bar{\psi}(z) \rangle_{c} \nonumber \\
&& + \langle \bar{\psi}_{\sigma}(x) \big(i\partial_{0}\gamma_{1} -
i\partial_{1}\gamma_{0} -
i\partial_{2}\gamma_{0}\gamma_{1}\gamma_{2}\big)
\psi_{\sigma}(x)\psi(y)\bar{\psi}(z) \rangle_{c} \nonumber \\
&=& \delta(x-y)\gamma_{0}\gamma_{1}G(x-z) + G(y-x)\gamma_{0}
\gamma_{1}\delta(x-z). \label{eq:currentgp01}
\end{eqnarray}
Apart from the current operators $j_{0}(x)$ and $j_{1}(x)$, here we
need to define one more current operator
\begin{eqnarray}
j_{012}(x) = \bar{\psi}_{\sigma}(x)\gamma_{012} \psi_{\sigma}(x),
\end{eqnarray}
where $\gamma_{012}=\gamma_{0}\gamma_{1}\gamma_{2}$. This new
current operator also corresponds to a new current vertex function
$\Gamma_{012}$, which is defined as
\begin{eqnarray}
\langle j_{012}(x)\psi(y)\bar{\psi}(z)\rangle_{c} &=& -\int d\xi
d\xi'G(y-\xi)\Gamma_{012}(\xi-x,x-\xi')G(\xi'-z),
\label{eq:currentvertex012}\\
\Gamma_{012}(\xi-x,x-\xi') &=& \int dq dp
e^{-i(p+q)\cdot(\xi-x)-ip\cdot(x-\xi')}\Gamma_{012}(q,p).
\label{eq:Fouriercurrentvertex012}.
\end{eqnarray}
The l.h.s. of Eq.~(\ref{eq:currentgp01}) is a little more
complicated than that of Eq.~(\ref{eq:currentgpI}). Originally, the
bilinear operators $j_{0}(x)$, $j_{1}(x)$, $j_{2}(x)$, and
$j_{012}(x)$ are defined as products of $\psi(x)$ and
$\bar{\psi}(x)$, which are supposed to be located at the same
time-space point $x$. In order to express the l.h.s. of
Eq.~(\ref{eq:currentgp01}) in terms of $j_{012}(x)$, we need to move
the partial derivative operator $\partial_{2}$ out of the mean
value. This can be achieved by employing the point-splitting
technique that is widely applied to regularize the short-distance
singularity caused by the locality of bilinear current operators in
high-energy physics \cite{Dirac, Schwinger, Jackiw, Takahashi78,
Peskin, Schnabl}. Using this technique \cite{Takahashi78, He01}, one
could re-define current operators at two very close but distinct
points $x$ and $x'$, namely
\begin{eqnarray}
j_{0,1,2,012}(x,x') = \bar{\psi}_{\sigma}(x') \gamma_{0,1,2,012}
\psi_{\sigma}(x).
\end{eqnarray}
The limit $x \rightarrow x'$ should be taken after all calculations
are completed. Now Eq.~(\ref{eq:currentgp01}) becomes
\begin{eqnarray}
&&i\partial_0\langle j_{1}(x)\psi(y)\bar{\psi}(z)\rangle_{c}
-i\partial_1\langle j_{0}(x)\psi(y)\bar{\psi}(z)\rangle_{c}
-\lim_{x'\to x}(i\partial_{2'}-i\partial_{2}) \langle
j_{012}(x,x')\psi(y)\bar{\psi}(z)\rangle_{c} \nonumber \\
&=&\delta(x-y)\gamma_{0}\gamma_{1}G(x-z) +
G(y-x)\gamma_{0}\gamma_{1}\delta(x-z).
\label{eq:divergencecurrent012}
\end{eqnarray}
Inserting Eq.~(\ref{eq:currentvertex1}) and
Eq.~(\ref{eq:currentvertex012}) into
Eq.~(\ref{eq:divergencecurrent012}) makes it possible to use
$\Gamma_{1}$, $\Gamma_{0}$, and $\Gamma_{012}$ to express the three
terms of l.h.s. of this equation, respectively. The first two terms
can be Fourier transformed in exactly the same way as
Eq.~(\ref{eq:Fouriersplitting1}). The third term is computed as
\begin{eqnarray}
&&\lim_{x'\to x}(i\partial_{2'}-i\partial_{2}) \langle
j_{012}(x,x')\psi(y)\bar{\psi}(z)\rangle_{c} \nonumber\\
&=& -\lim_{x'\to x}(i\partial_{2'}-i\partial_{2}) \int d\xi
d\xi'G(y-\xi)\Gamma_{012}(\xi-x',x-\xi')G(\xi'-z) \nonumber \\
&=&-\lim_{x'\to x}(i\partial_{2'}-i\partial_{2}) \int d\xi d\xi'
\int dp dq dp'dq'e^{-i(p+q)\cdot (y-\xi)}G(p+q) \nonumber \\
&& \times e^{-i(p'+q')\cdot(\xi-x')-ip'\cdot(x-\xi')}
\Gamma_{012}(q',p')e^{-ip\cdot(\xi'-z)}G(p) \nonumber \\
&=&-\lim_{x'\to x}(i\partial_{2'}-i\partial_{2}) \int dp dq
dp'dq'e^{-i(p+q)\cdot y} G(p+q)\delta\big(p+q-(p'+q')\big) \nonumber
\\
&& \times e^{i(p'+q')\cdot x' - ip'\cdot x}\Gamma_{012}(q',p')
\delta(p'-p)e^{ip\cdot z}G(p) \nonumber \\
&=& -\lim_{x'\to x}(i\partial_{2'}-i\partial_{2}) \int dp dq
e^{-i(p+q)\cdot (y-x')}e^{-ip\cdot (x-z)}G(p+q)
\Gamma_{012}(q,p)G(p) \nonumber \\
&=& \lim_{x'\to x} \int dp dq (p_2+q_2+p_2)e^{-i(p+q)\cdot(y-x')}
e^{-ip\cdot (x-z)}G(p+q)\Gamma_{012}(q,p)G(p) \nonumber \\
&=& \int dpdqe^{-i(p+q)\cdot(y-x)}e^{-ip\cdot (x-z)}(2p_2+q_2)
G(p+q)\Gamma_{012}(q,p)G(p).
\end{eqnarray}
Finally, we obtain from Eq.~(\ref{eq:divergencecurrent012}) the
following identity
\begin{eqnarray}
q_{0}\Gamma_{1}(q,p)-q_{1}\Gamma_{0}(q,p)-(2p_{2}+q_{2})\Gamma_{012}(q,p)
= -G^{-1}(p+q)\gamma_{01} - \gamma_{01}G^{-1}(p). \label{eq:WTI2}
\end{eqnarray}
This identity has an analogous form to the ordinary WTI given by
Eq.~(\ref{eq:WTI1}). There is an important different between them.
The ordinary WTI is induced by the U(1)-symmetry of the Lagrangian
density. In contrast, the identity given by Eq.~(\ref{eq:WTI2})
originates from the invariance of the partition function under the
transformation $\psi \rightarrow e^{i\theta\gamma_{01}}\psi$, which
is not a symmetry of the model as it apparently changes the
Lagrangian density.

Thus far, we have derived two identities obeyed by four different
current vertex functions $\Gamma_{0}(q,p)$, $\Gamma_{1}(q,p)$,
$\Gamma_{2}(q,p)$, and $\Gamma_{012}(q,p)$. We still need at least
two more identities to completely determine each of these functions.
For $\gamma_{m}=\gamma_{02}$, the identity of
Eq.~(\ref{eq:currentgp}) becomes
\begin{eqnarray}
&&\langle\big(i\partial_0\bar{\psi}_{\sigma} \gamma_{2} +
i\partial_{1}\bar{\psi}_{\sigma}\gamma_{0}\gamma_{1}\gamma_{2} -
i\partial_2\bar{\psi}_{\sigma}\gamma_{0}\big)\psi_{\sigma}
\psi(y)\bar{\psi}(z) \rangle_{c} \nonumber \\
&& +\langle\bar{\psi}_{\sigma}\big(i\partial_{0}\gamma_{2} -
i\partial_{1}\gamma_{0}\gamma_{1}\gamma_{2}-i\partial_{2}\gamma_{0}\big)
\psi_{\sigma}\psi(y)\bar{\psi}(z) \rangle_{c} \nonumber \\
&=& \delta(x-y)\gamma_{0}\gamma_{2}G(x-z) + G(y-x)\gamma_{0}
\gamma_{2}\delta(x-z).\label{eq:currentgp02}
\end{eqnarray}
Applying point-splitting trick to this equation gives rise to
\begin{eqnarray}
&&i\partial_{0}\langle j_{2}(x)\psi(y)\bar{\psi}(z)\rangle_{c} +
\lim_{x'\to x}(i\partial_{1'}-i\partial_{1})\langle
j_{012}(x,x')\psi(y)\bar{\psi}(z)\rangle_{c} -i\partial_{2}\langle
j_{0}(x)\psi(y)\bar{\psi}(z)\rangle_{c} \nonumber \\
&=& \delta(x-y)\gamma_{0}\gamma_{2}G(x-z) +
G(y-x)\gamma_{0}\gamma_{2}\delta(x-z). \label{eq:pointsplitted3}
\end{eqnarray}
Finally, choosing $\gamma_{m}=i\gamma_{12}$ makes
Eq.~(\ref{eq:currentgp}) to become
\begin{eqnarray}
&&\langle\big(i\partial_0\bar{\psi}_{\sigma}
i\gamma_{0}\gamma_{1}\gamma_{2} + i\partial_{1}\bar{\psi}_{\sigma}
i\gamma_{2} - i\partial_2\bar{\psi}_{\sigma}i\gamma_{1}\big)
\psi_{\sigma}\psi(y)\bar{\psi}(z)\rangle_{c}
\nonumber\\&&
+ \langle\bar{\psi}_{\sigma}\big(i\partial_{0}
i\gamma_{0}\gamma_{1}\gamma_{2} -
i\partial_{1}i\gamma_{2}-i\partial_{2}(-i\gamma_{1})\big)
\psi_{\sigma}\psi(y)\bar{\psi}(z) \rangle_{c} \nonumber \\
&=& \delta(x-y)i\gamma_{1}\gamma_{2}G(x-z) -
G(y-x)i\gamma_{1}\gamma_{2}\delta(x-z),\label{eq:currentgpi12}
\end{eqnarray}
which can be re-written as
\begin{eqnarray}
&&i\partial_0\langle j_{012}(x)\psi(y)\bar{\psi}(z)\rangle_{c}
+\lim_{x'\to x}(i\partial_{1'}-i\partial_{1})\langle
j_{2}(x,x')\psi(y)\bar{\psi}(z)\rangle_{c} \nonumber \\
&& -\lim_{x'\to x}(i\partial_{2'}-i\partial_{2})\langle
j_{1}(x,x')\psi(y)\bar{\psi}(z)\rangle_{c}\nonumber \\
&=&\delta(x-y)\gamma_{1}\gamma_{2}G(x-z) -
G(y-x)\gamma_{1}\gamma_{2}\delta(x-z).
\label{eq:pointsplitted4}
\end{eqnarray}
After performing Fourier transformations, we find that
Eq.~(\ref{eq:pointsplitted3}) and Eq.~(\ref{eq:pointsplitted4})
yield two identities:
\begin{eqnarray}
q_{0}\Gamma_{2}(q,p)-q_{2}\Gamma_{0}(q,p)+(2p_{1}+q_{1})\Gamma_{012}(q,p)
&=& -G^{-1}(p+q)\gamma_{02} - \gamma_{02}G^{-1}(p), \label{eq:WTI3} \\
q_{0}\Gamma_{012}(q,p)-(2p_{2}+q_{2})\Gamma_{1}(q,p) +
(2p_{1}+q_{1})\Gamma_{2}(q,p) &=& G^{-1}(p+q)\gamma_{12} -
\gamma_{12}G^{-1}(p). \label{eq:WTI4}
\end{eqnarray}

The four independent identities given by Eq.~(\ref{eq:WTI1}),
Eq.~(\ref{eq:WTI2}), Eq.~(\ref{eq:WTI3}), and Eq.~(\ref{eq:WTI4})
are generated respectively by making the following four
infinitesimal transformations of the spinor field:
$$\psi \to e^{i\theta}\psi, \quad \psi \to e^{i\theta \gamma_{01}}
\psi, \quad \psi \to e^{i\theta \gamma_{02}}\psi, \quad \psi \to
e^{-\theta\gamma_{12}}\psi.$$ Among these transformations, the first
one keeps the Lagrangian density intact and thus Eq.~(\ref{eq:WTI1})
is a genuine symmetry-induced WTI. The rest three transformations
are clearly not symmetries of the model. The forth one is not even a
unitary transformation. Therefore, the last three identities are
different from Eq.~(\ref{eq:WTI1}). Nevertheless, we would regard
all of the four identities as generalized WTIs for two reasons.
First, they have very similar forms. Second, they can be derived in
a unified way from the invariance of the partition function.

These four generalized WTIs can be expressed in the following
compact form:
\begin{eqnarray}
M \begin{pmatrix}\Gamma_0(q,p)\\ \Gamma_1(q,p)\\ \Gamma_2(q,p)\\
\Gamma_{012}(q,p)
\end{pmatrix}=
\begin{pmatrix}
G^{-1}(p+q)-G^{-1}(p)\\
-G^{-1}(p+q)\gamma_{01}-\gamma_{01}G^{-1}(p)\\
-G^{-1}(p+q)\gamma_{02}-\gamma_{02}G^{-1}(p)\\
G^{-1}(p+q)\gamma_{12}-\gamma_{12}G^{-1}(p)
\end{pmatrix}.
\label{eq:fourwtis}
\end{eqnarray}
Here the matrix $M$ is given by
\begin{eqnarray}
M = \begin{pmatrix}
q_0&-q_1&-q_2&0\\
-q_1&q_0&0&-(2p_2+q_2)\\
-q_2&0&q_0&+(2p_1+q_1)\\
0&-(2p_2+q_2)&(2p_1+q_1)&q_0
\end{pmatrix}.
\end{eqnarray}
Now each of the four unknown functions $\Gamma_{0}(q,p)$,
$\Gamma_{1}(q,p)$, $\Gamma_{2}(q,p)$, and $\Gamma_{012}(q,p)$ can be
determined by solving the four coupled identities shown in
Eq.~(\ref{eq:fourwtis}). According to Eq.~(\ref{eq:dsefinal}), we
only need to know $\Gamma_0(q,p)$. From Eq.~(\ref{eq:fourwtis}), it
is easy to obtain
\begin{eqnarray}
\Gamma_{0}(q,p) &=& \frac{1}{|M|}\Big[|M_{11}|
\left(G^{-1}(p+q)\gamma_0 - \gamma_0G^{-1}(p)\right) -
|M_{21}|\left(-G^{-1}(p+q)\gamma_{01} - \gamma_{01}G^{-1}(p)\right)
\nonumber \\
&& +|M_{31}|\left(-G^{-1}(p+q)\gamma_{02} -
\gamma_{02}G^{-1}(p)\right) -|M_{41}|\left(G^{-1}(p+q)\gamma_{012} -
\gamma_{012}G^{-1}(p)\right)\Big],\label{eq:gamma0}
\end{eqnarray}
where
\begin{eqnarray}
|M| &=& q_{0}^{2}\left(q_{0}^{2}-q_{1}^{2}-q_{2}^{2} -
(2p_{1}+q_{1})^{2}-(2p_{2}+q_{2})^{2}\right) +
\big(q_1(2p_1+q_1)+q_2(2p_2+q_2)\big)^{2}, \nonumber \\
|M_{11}|&=&q_0\left(q_{0}^{2}-(2p_{1}+q_{1})^{2} -
(2p_2+q_{2})^{2}\right), \nonumber \\
|M_{21}| &=& q_1(2p_1+q_1)^{2}-q_1 q_{0}^{2} + q_2(2p_1 + q_1)
(2p_2+q_2), \nonumber \\
|M_{31}| &=& q_2 q_0^2 - q_2(2p_2+q_2)^2 - q_1(2p_1+q_1)(2p_2+q_2),
\nonumber \\
|M_{41}| &=& q_0q_1(2p_2+q_2)-q_0q_2(2p_1+q_1).
\end{eqnarray}
Since $\Gamma_0(q,p)$ depends only on the full fermion propagator
$G(p)$, the DS equation of $G(p)$ given by Eq.~(\ref{eq:dsefinal})
becomes completely self-closed and can be solved by the iteration
method \cite{Liu21}. In passing, we have already confirmed that
$\Gamma_{0}(q,p)$ does not exhibit any singularity since the zeroes
of the denominator and numerator cancel each other out.

\section{Numerical results of renormalized velocity \label{Sec:numerics}}

In this section, we discuss the physical implications of the
numerical results of Eq.~(\ref{eq:dsefinal}). It appears to be more
convenient to perform numerical calculations if the Matsubara
formalism of finite-temperature field theory is adopted. The real
time $t$ appearing in the DS equation of fermion propagator should
be replaced with the Matsubara time $\tau$, where $\tau \in [-T,T]$.
The fermion momentum $p=(p_{0},\mathbf{p})$ becomes
$p=(i\epsilon_n,\mathbf{p})$, where $i\epsilon_n = i(2n+1)\pi T$,
and the boson momentum $q = (q_{0},\mathbf{q})$ becomes
$q=(i\omega_{n^{\prime}},\mathbf{q})$, where $i\omega_{n^{\prime}} =
i 2n^{\prime}\pi T$. $n$ and $n'$ take all the integers.

As shown by Eq.~(\ref{eq:dsefinal}), the DS equation of $G(p)$
contains the free propagators of two bosons. The free phonon
propagator is
\begin{eqnarray}
D_{0}(q) = \frac{2\Omega_{\mathbf{q}}}{(i\omega_{n'})^2 -
\Omega_{\mathbf{q}}^2},\label{eq:freedq}
\end{eqnarray}
where the phonon dispersion is $\Omega_{\mathbf{q}} =
c_{s}|\mathbf{q}|$ with $c_s$ being the phonon velocity
\cite{Roy14}. The EPI strength parameter $g$ is a function of phonon
momentum and formally defined \cite{Roy14} as
\begin{eqnarray}
g \equiv g(q) = \sqrt{\lambda q/c_{s}},\label{eq:gq}
\end{eqnarray}
where $q=|\mathbf{q}|$ is phonon momentum and $\lambda$ is a
dimensionless tuning parameter. The precise value of $\lambda$ in
undoped graphene is material dependent and should be determined by
performing careful first-principle calculations. Here we regard
$\lambda$ as a freely varying parameter and make a generic
(material-independent) analysis. The free propagator of $A$ boson is
\begin{eqnarray}
F_{0}(q) = \frac{2\pi \alpha}{|\mathbf{q}|},\label{eq:freefq}
\end{eqnarray}
which has the same form as the bare Coulomb interaction function.
The fine structure constant
\begin{eqnarray}
\alpha = \frac{e^{2}}{v_{F}\varepsilon},\label{eq:alpha}
\end{eqnarray}
characterizes the effective strength of Coulomb interaction
\cite{Castroneto09, Sarma11, Kotov12}. It is well-known that $\alpha
= 0.8$ for graphene on SiO$_{2}$ substrate and $\alpha=2.2$ for
graphene suspended in vacuum.

After incorporating the corrections induced by interactions, the
free boson propagators will become dressed. The renormalization of
such model parameters as $c_{s}$ and $\varepsilon$ can be studied by
comparing the dressed boson propagators with the free boson
propagators. In the literature (see Ref.~\cite{Kotov12} for a
review), the dressed boson propagators are usually calculated by
employing the random phase approximation (RPA). In undoped graphene,
the RPA-level, one-loop polarization function is found \cite{Son07,
Kotov12} to have the form $\Pi_{\mathrm{RPA}}(q) = -\frac{N}{8}
\frac{\mathbf{q}^{2}}{\sqrt{q_{0}^{2} + v^{2}\mathbf{q}^{2}}}$. Then
the dressed phonon propagator is $D_{\mathrm{RPA}}(q) =
\frac{1}{D_{0}^{-1}(q)+\Pi_{\mathrm{RPA}}(q)}$ and the dressed $A$
boson propagator (i.e., renormalized Coulomb interaction) is
$F_{\mathrm{RPA}}(q) = \frac{1}{F_{0}^{-1}(q) +
\Pi_{\mathrm{RPA}}(q)}$. Now both $D_{\mathrm{RPA}}(q)$ and
$D_{\mathrm{RPA}}(q)$ are proportional to $\sim 1/N$. This provides
a basis to classify all the Feynman diagrams according to the powers
of $1/N$. The $1/N$ expansion has been adopted to investigate the
physical effects of the Coulomb interaction in both perturbative
calculations \cite{Son07, Son08, Hofmann14, Kotov12} and
non-perturbative DS equation studies \cite{Khveshchenko01, Gorbar02,
Khveshchenko04, Khveshchenko09, Liu09, Gamayun10, WangLiu12,
WangLiu14}. However, $1/N$ expansion is well justified only in the
$N\rightarrow \infty$ limit. Given that the physical flavor is
rather small ($N=2$), the validity of $1/N$ expansion is in doubt.

Using our approach, the DS equations of fermion and boson
propagators are decoupled \cite{Liu21, Pan21}. Thus the
renormalization of boson propagators should be treated in a very
different way from previous perturbative and non-perturbative
calculations. Notice that the DS equation of full fermion propagator
$G(p)$, given by Eq.~(\ref{eq:dsefinal}), depends on the free boson
propagators $D_{0}(q)$ and $F_{0}(q)$, rather than the dressed boson
propagators $D(q)$ and $F(q)$. The interaction effects on the bosons
are already indirectly embodied in the current vertex function
$\Gamma_{0}(q,p)$. There would be an incorrect double counting if
the dressed boson propagators $D(q)$ and $F(q)$ are substituted into
Eq.~(\ref{eq:dsefinal}). Therefore, the parameters $c_{s}$ and
$\varepsilon$ appearing in $D_{0}(q)$ and $F_{0}(q)$ should take
their bare values and must not be renormalized. For similar reasons,
we need to use the bare value of EPI strength parameter $g$, whose
renormalization is already taken into account by the function
$\Gamma_{0}(q,p)$. The electric charge $e$ is also not renormalized
\cite{Ye98, Herbut06, Pan21}. Different from the above parameters,
the fermion velocity $v_{F}$ is renormalized by interactions. Below
we demonstrate how to obtain the renormalized fermion velocity based
on the solutions of $G(p)$.

The free fermion propagator is
\begin{eqnarray}
G_{0}(p) = \frac{1}{i\epsilon_n\gamma_0 - \mathbf{\gamma}\cdot
\mathbf{p}} =-\frac{i\epsilon_n\gamma_0 + \mathbf{\gamma}\cdot
\mathbf{p}} {\epsilon_n^2+\mathbf{p}^2}.\label{eq:g0p}
\end{eqnarray}
Incorporating the interaction effects turns this free propagator
into a full propagator that can be expressed as
\begin{eqnarray}
G(p) = \frac{1}{A_0(\epsilon_{n},\mathbf{p})i\epsilon_n\gamma_0 -
A_1(\epsilon_{n},\mathbf{p})\mathbf{\gamma}\cdot \mathbf{p}} =
-\frac{A_0(\epsilon_{n},\mathbf{p})i\epsilon_n\gamma_0 +
A_1(\epsilon_{n},\mathbf{p})\mathbf{\gamma}\cdot \mathbf{p}}
{A_0^2(\epsilon_{n},\mathbf{p})\epsilon_{n}^2
+A_1^2(\epsilon_{n},\mathbf{p})\mathbf{p}^2}.
\end{eqnarray}
The interactions effects are embodied in the two renormalization
functions $A_{0}(\epsilon_{n},\mathbf{p})$ and
$A_{1}(\epsilon_{n},\mathbf{p})$. Inserting $D_{0}(q)$, $F_{0}(q)$,
$G_{0}(p)$, and $G(p)$ together with the function $\Gamma_{0}(q,p)$
given by Eq.~(\ref{eq:gamma0}) into Eq.~(\ref{eq:dsefinal}) yields
two self-consistent integral equations of
$A_{0}(\epsilon_{n},\mathbf{p})$ and
$A_{1}(\epsilon_{n},\mathbf{p})$.

For readers' convenience, below we list all the formulae needed to
express the self-closed DS equation of the full fermion propagator:
\begin{eqnarray}
G^{-1}(\epsilon_n,\mathbf{p}) &=& i\epsilon_n\gamma_0 -
\mathbf{\gamma\cdot p} +T\sum_{n'}\int\frac{d^2\mathbf{q}}{(2\pi)^2}
\left[g^2D_0(q)+F_0(q)\right] G(p+q)\Gamma_0(q,p),\\
\Gamma_0(q,p) &=& \frac1{|M|}\Big[|M_{11}|\left(G^{-1}(p+q) \gamma_0
-\gamma_0 G^{-1}(p)\right)-|M_{21}|\left(-G^{-1}(p+q)\gamma_{01} -
\gamma_{01} G^{-1}(p)\right) \nonumber \\
&& +|M_{31}|\left(-G^{-1}(p+q)\gamma_{02} -
\gamma_{02}G^{-1}(p)\right) -|M_{41}|\left(G^{-1}(p+q)\gamma_{012} -
\gamma_{012}G^{-1}(p)\right)\Big],\\
|M| &=& \omega_{n'}^{2} \left(\omega_{n'}^{2} + q_{1}^{2} +
q_{2}^{2} + (2p_1+q_1)^2+(2p_2+q_2)^2\right)
\nonumber \\
&& +\big(q_1(2p_1+q_1)+q_2(2p_2+q_2)\big)^2,\label{Eq:|M|} \\
|M_{11}| &=& -i\omega_{n'}\left(\omega_{n'}^{2}+(2p_1 +
q_1)^{2}+(2p_{2}+q_{2})^{2}\right)\label{Eq:M11}, \\
|M_{21}| &=& q_1\left(\omega_{n'}^{2} +
(2p_{1}+q_{1})^{2}\right)+q_2(2p_1+q_1)(2p_2+q_2)\label{Eq:M21}, \\
|M_{31}|&=&-q_2\left(\omega_{n'}^2+(2p_2+q_2)^2\right) -
q_1(2p_1+q_1)(2p_2+q_2)\label{Eq:M31}, \\
|M_{41}|&=&-i\omega_{n'} q_2(2p_1+q_1)+i\omega_{n'}q_1(2p_2+q_2)\label{Eq:M41}.
\end{eqnarray}
To facilitate numerical computations, we re-write these equations in
the polar coordinate. We select $\mathbf{p}$ as the polar axis and
define a new momentum $k=p+q$. Then $k_1 = |\mathbf{k}|\cos\theta$,
$k_{2} = |\mathbf{k}|\sin\theta$, $p_1=|\mathbf{p}|$, and $p_2=0$.
Then Eqs.~(\ref{Eq:|M|}-\ref{Eq:M41}) become
\begin{eqnarray}
|M|&=&\omega_{n'}^{2}\left(\omega_{n'}^2 +
2|\mathbf{k}|^2+2|\mathbf{p}|^2\right) +
\left(|\mathbf{k}|^2-|\mathbf{p}|^2\right)^2, \\
|M_{11}|&=&-i\omega_{n'}\left(\omega_{n'}^2 +
|\mathbf{k}|^2+|\mathbf{p}|^2
+ 2|\mathbf{k}||\mathbf{p}|\cos\theta\right), \\
|M_{21}|&=&-\left(\omega_{n'}^2+|\mathbf{p}|^2-|\mathbf{k}|^2\right)
|\mathbf{p}|+\left(\omega_{n'}^2+|\mathbf{k}|^2-|\mathbf{p}|^2\right)
|\mathbf{k}|\cos\theta, \\
|M_{31}|&=&-|\mathbf{k}|\sin\theta \left(\omega_{n'}^2 +
|\mathbf{k}|^2-|\mathbf{p}|^2\right), \\
|M_{41}| &=& -i2\omega_{n'}|\mathbf{k}|\sin\theta|\mathbf{p}|.
\end{eqnarray}
The self-consistent integral equations of $A_0(p)$ and $A_1(p)$ are
given by
\begin{eqnarray}
A_0(p)\epsilon_n &=&\epsilon_n + T\sum_{n'} \int
\frac{|\mathbf{k}|d|\mathbf{k}|}{2\pi} \frac{1}{A_0^2(k)
\epsilon_{n'+n}^{2} + A_{1}^{2}(k)|\mathbf{k}|^2} \nonumber \\
&& \times \Big[A_0(k)\epsilon_{n'+n}\Big(f_{0kp1}\big(A_0(k)
\epsilon_{n'+n}-A_0(p)\epsilon_n\big) + f_{k0p1}A_1(k)|\mathbf{k}| -
f_{k1p0}A_1(p)|\mathbf{p}|\Big)\nonumber \\
&& +A_1(k)|\mathbf{k}|\Big(-f_{k0p1}\big(A_0(k)\epsilon_{n'+n} -
A_0(p)\epsilon_n\big) + f_{0kp1}A_1(k)|\mathbf{k}| -
f_{1kp0}A_1(p)|\mathbf{p}|\Big)\Big], \label{eq:eqa0} \\
A_1(p)|\mathbf{p}| &=& |\mathbf{p}| - T\sum_{n'}\int
\frac{|\mathbf{k}|d|\mathbf{k}|}{2\pi}\frac{1}{A_0^2(k)
\epsilon_{n'+n}^2+A_1^2(k)|\mathbf{k}|^{2}} \nonumber \\
&&\times \Big[A_0(k)\epsilon_{n'+n}\Big(f_{k1p0}\big(A_0(k)
\epsilon_{n'+n} - A_0(p)\epsilon_n\big)-f_{1kp0}A_1(k) |\mathbf{k}|
+ f_{0kp1}A_1(p)|\mathbf{p}|\Big) \nonumber \\
&& +A_1(k)|\mathbf{k}|\Big(f_{1kp0}\big(A_0(k)\epsilon_{n'+n} -
A_0(p)\epsilon_{n}\big) + f_{k1p0}A_1(k)|\mathbf{k}| -
f_{k0p1}A_1(p)|\mathbf{p}|\Big)\Big]. \label{eq:eqa1}
\end{eqnarray}
Here, we have defined several quantities:
\begin{eqnarray}
\Sigma_{0} &=& -\int\frac{d\theta}{2\pi} \left(g^{2}D_{0} +
F_{0}\right), \\
\Sigma_{1} &=& -\int\frac{d\theta}{2\pi}
\left(g^{2}D_{0}+F_{0}\right)\cos\theta, \\
f_{k0p1} &=& \frac{1}{|M|}\left[\omega_{n'}^{2}
\left(|\mathbf{k}| \Sigma_0-|\mathbf{p}|\Sigma_1\right) +
\left(|\mathbf{k}|^2-|\mathbf{p}|^2\right)
\left(|\mathbf{k}|\Sigma_0+|\mathbf{p}|\Sigma_1\right)\right],
\\
f_{k1p0} &=& \frac{1}{|M|} \left[\omega_{n'}^{2}
\left(|\mathbf{k}| \Sigma_{1}-|\mathbf{p}|\Sigma_0\right) +
\left(|\mathbf{k}|^{2}-|\mathbf{p}|^2\right)
\left(|\mathbf{k}|\Sigma_{1}+|\mathbf{p}|\Sigma_0\right)\right], \\
f_{0kp1} &=& \frac{\omega_{n'}}{|M|} \left[\left(
\omega_{n^{\prime}}^{2} + |\mathbf{k}|^{2} + |\mathbf{p}|^2\right)
\Sigma_{0}+2|\mathbf{k}||\mathbf{p}|\Sigma_1\right],
\\
f_{1kp0} &=& \frac{\omega_{n^{\prime}}}{|M|}
\left[\left(\omega_{n'}^{2} + |\mathbf{k}|^{2} +
|\mathbf{p}|^2\right)\Sigma_1 +
2|\mathbf{k}||\mathbf{p}|\Sigma_0\right].
\end{eqnarray}

We have solved Eqs.~(\ref{eq:eqa0}-\ref{eq:eqa1}) by means of the
iteration method \cite{Liu21}. Since we are mainly interested in the
zero-$T$ behavior of fermion velocity renormalization, we take the
limit $T\rightarrow 0$. The energy-momentum dependence of
renormalized fermion velocity is computed from the ratio:
\begin{eqnarray}
v(\epsilon_{n},\mathbf{p}) = \frac{A_{1}(\epsilon_{n},
\mathbf{p})}{A_{0}(\epsilon_{n},\mathbf{p})}.
\end{eqnarray}

\begin{figure}[htbp]
\centering \subfigure[]{\label{Fig.sub.1}
\includegraphics[width=1.96in]{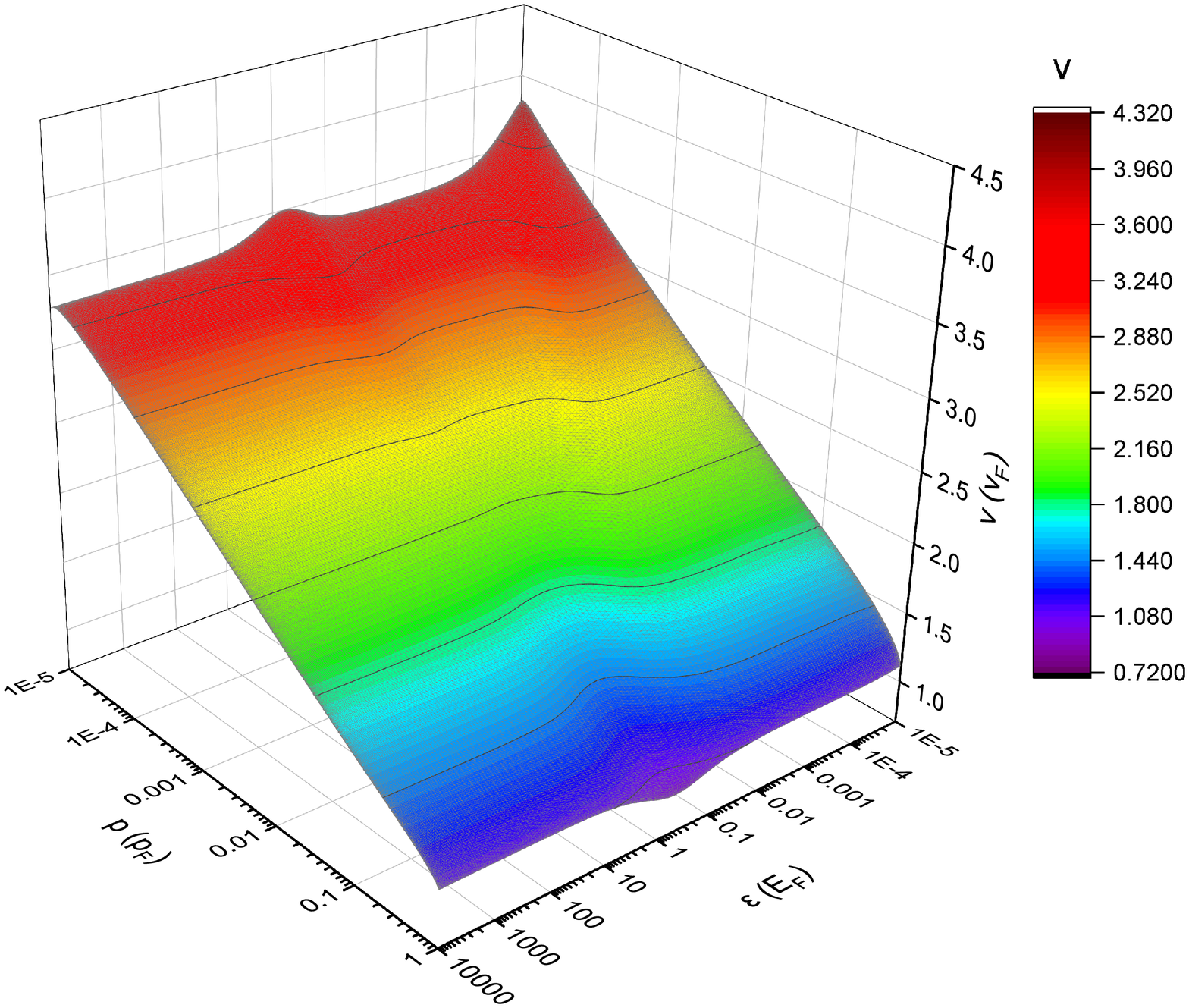}}
\subfigure[]{\label{Fig.sub.2}
\includegraphics[width=1.96in]{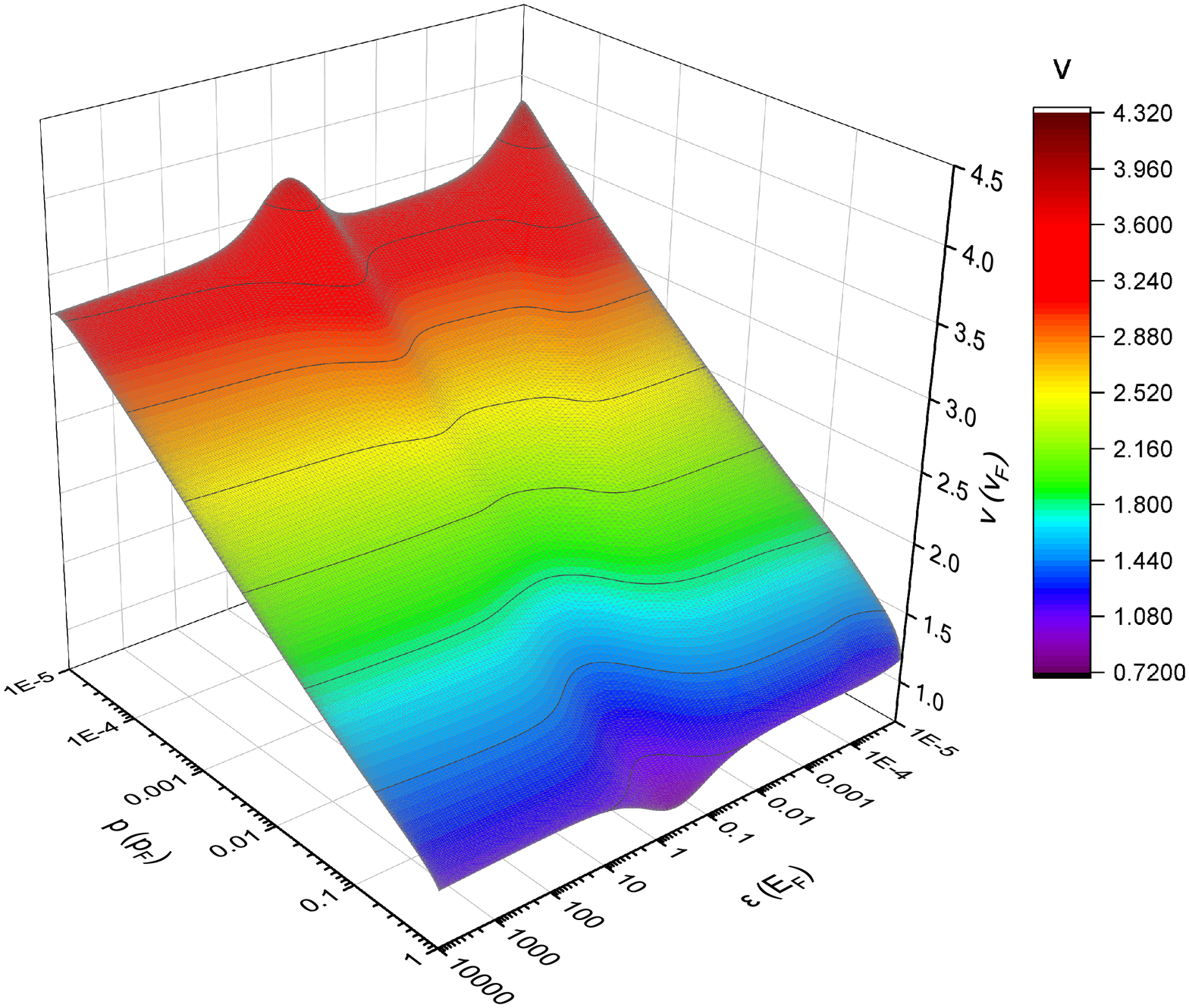}}
\subfigure[]{\label{Fig.sub.3}
\includegraphics[width=1.96in]{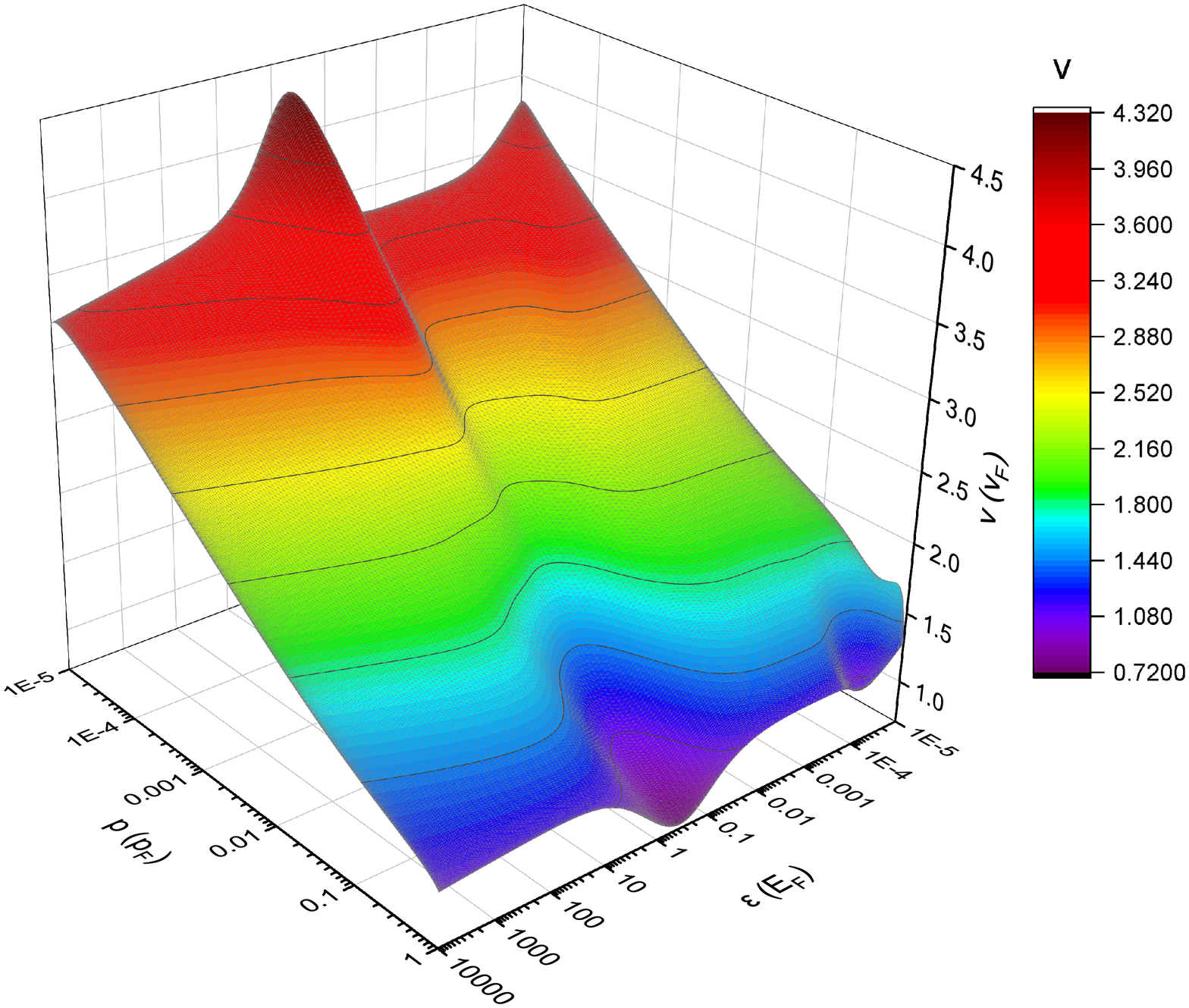}}
\subfigure[]{\label{Fig.sub.4}
\includegraphics[width=1.96in]{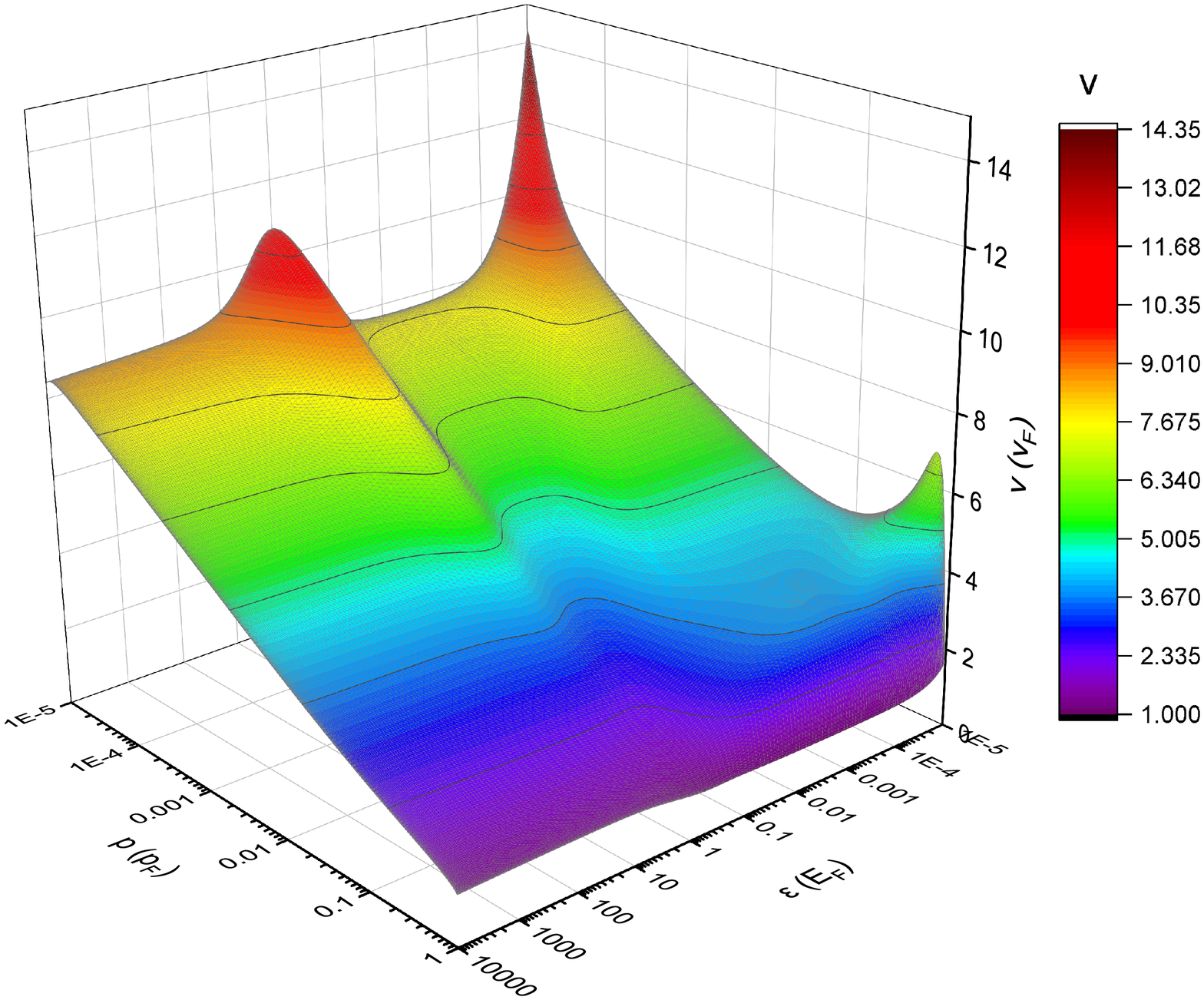}}
\subfigure[]{\label{Fig.sub.5}
\includegraphics[width=1.96in]{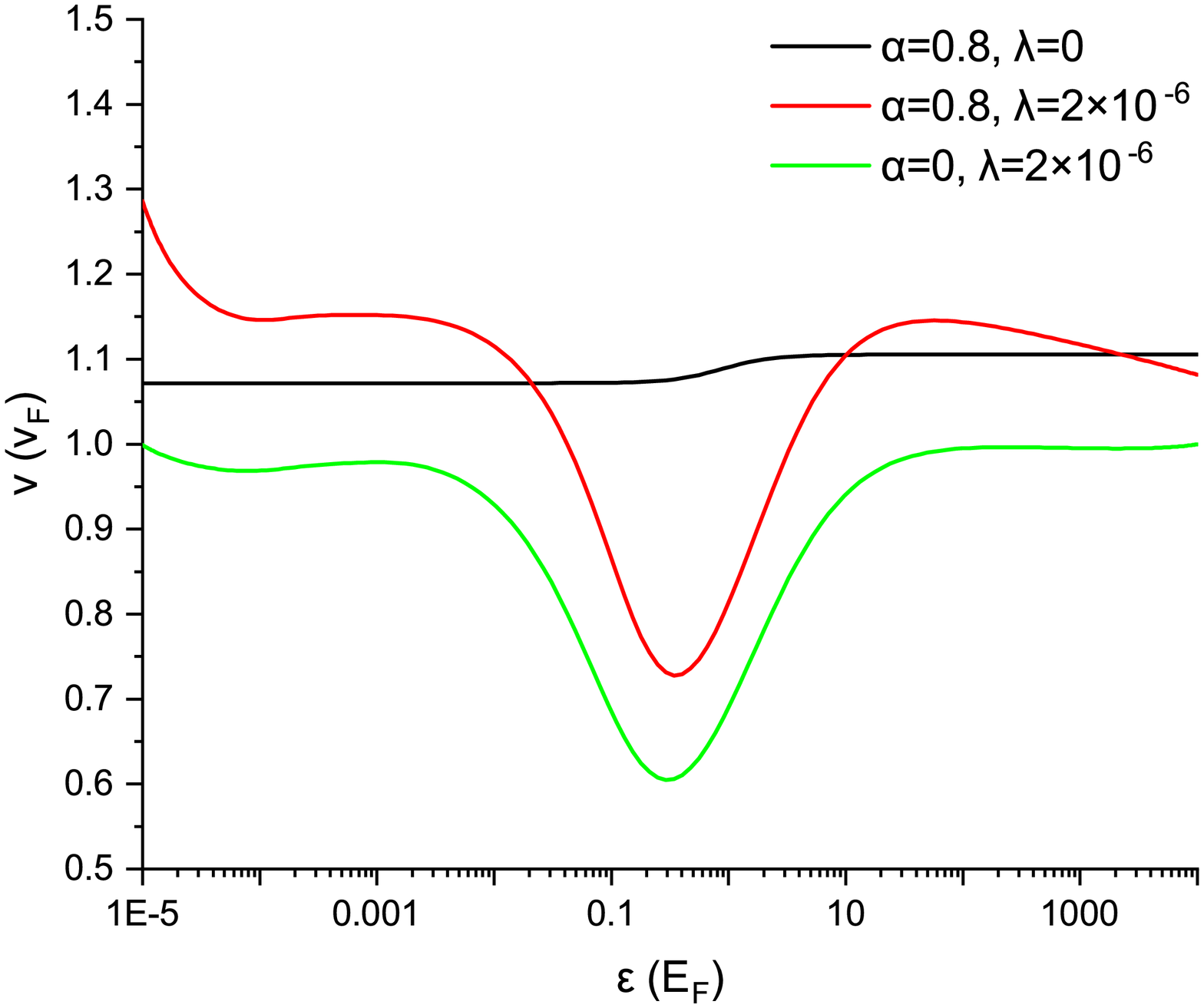}}
\subfigure[]{\label{Fig.sub.6}
\includegraphics[width=1.96in]{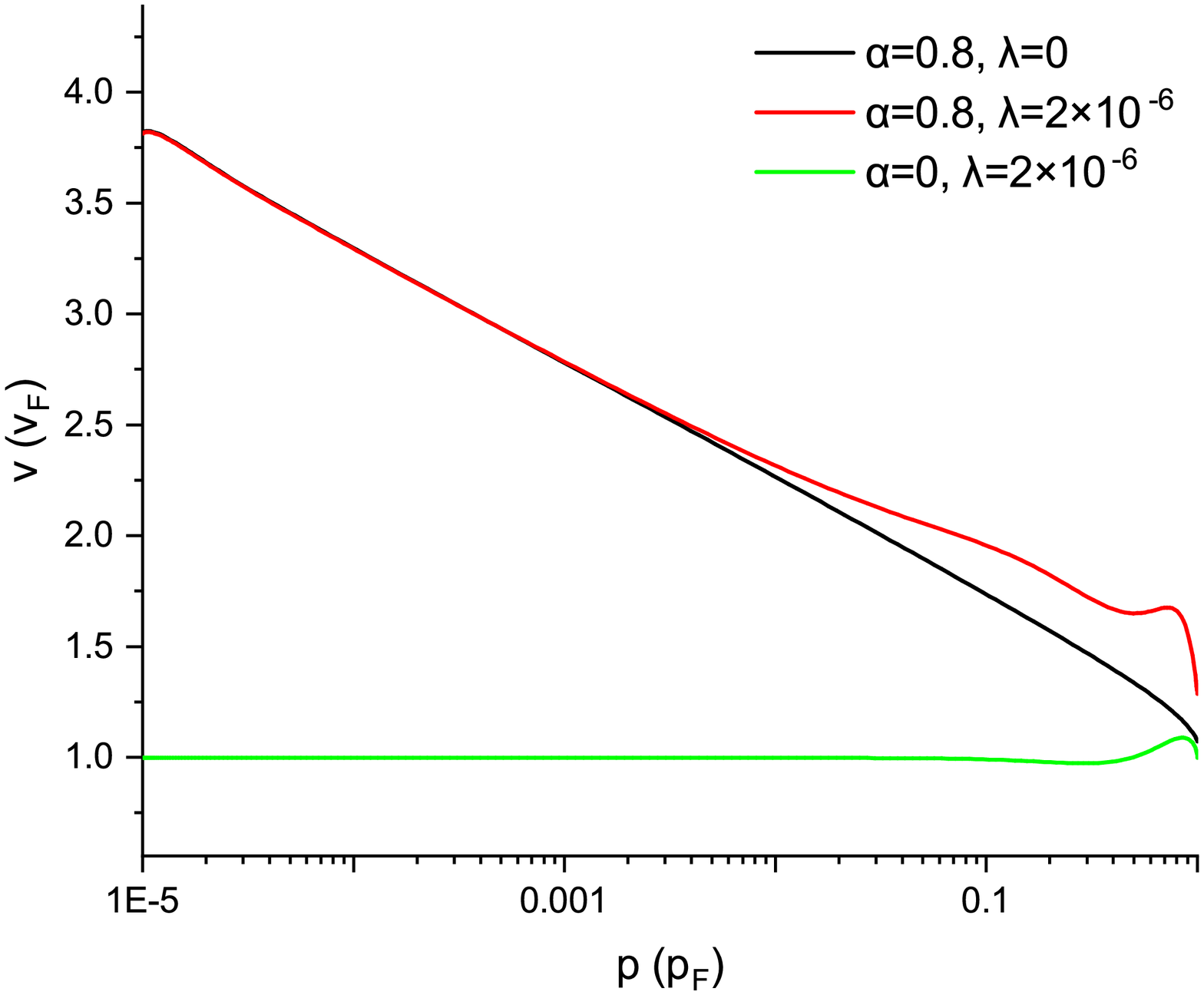}}
\caption{Full energy-momentum dependence of renormalized fermion
velocity $v$ are presented in (a-d). Phonon velocity is fixed at
$c_{s} = 1\times 10^{-4}$ (in unit of bare fermion velocity
$v_{F}$). (a) $\lambda=0.5\times 10^{-6}$, $\alpha=0.8$. (b)
$\lambda=1\times 10^{-6}$, $\alpha=0.8$. (c)
$\lambda=2\times10^{-6}$, $\alpha=0.8$. (d) $\lambda=1\times
10^{-6}$, $\alpha=2.2$. (e) Energy-dependence of $v$ as
$|\mathbf{p}|\rightarrow p_{F}$. (f) Momentum-dependence of $v$ as
$\epsilon \rightarrow 0$.} \label{fig:velocity}
\end{figure}

The numerical results of $v(\epsilon_{n},\mathbf{p})$ are plotted in
Fig.~\ref{fig:velocity}. The energy (momentum) is in unit of Fermi
energy $E_{F}$ (Fermi momentum $p_{F}$). One observes from
Fig.~\ref{fig:velocity}(a-c) that $v(\epsilon_{n},\mathbf{p})$
exhibits a clear non-monotonic dependence on energy for any fixed
$|\mathbf{p}|$ due to the interplay of two interactions. In
comparison, as shown in Ref.~\cite{Pan21}, $v$ is energy independent
if we only consider the Coulomb interaction. Thus the non-monotonic
energy-dependence of $v$ is dominantly induced by EPI. To see this
fact more explicitly, we plot $v(\epsilon)$ in
Fig.~\ref{fig:velocity}(e) in the $|\mathbf{p}|\rightarrow p_{F}$
limit. As the EPI strength parameter $\lambda$ increases, the
non-monotonicity becomes more pronounced, which can be seen by
comparing the results shown in Fig.~\ref{fig:velocity}(a-c).

Moreover, we find that $v$ is a decreasing function of
$|\mathbf{p}|$ at any fixed energy for $\alpha=0.8$, no matter
whether the effects of EPI are taken into account. For $\alpha=2.2$,
$v$ first decreases with growing $|\mathbf{p}|$, but tends to
increase as $|\mathbf{p}|$ approaches its ultraviolet cutoff. This
upturn behavior is shown in Fig.~\ref{fig:velocity}(d). According to
Fig.~\ref{fig:velocity}(f), EPI makes little contribution to the
$|\mathbf{p}|$-dependence of $v$. In particular, adding EPI to the
system does not change the logarithmic $|\mathbf{p}|$-dependence of
$v(\mathbf{p})$ in the small-$|\mathbf{p}|$ region caused purely by
the Coulomb interaction. This result provides a natural explanation
of the surprisingly good agreement between the experimental result
of $v(\mathbf{p})$ measured in realistic graphene materials
\cite{Elias11, Lanzara11, Chae12} and the theoretical result of
$v(\mathbf{p})$ calculated without taking into account the impact of
EPI \cite{Gonzalez94, Pan21}. We see from
Fig.~\ref{fig:velocity}(a-d) that the renormalized velocity $v$
seems to increase abruptly if $\epsilon \to 0$ and
$|\mathbf{p}|\rightarrow 0$. As discussed in Ref.~\cite{Pan21}, this
is an artifact caused by infrared cutoffs and the logarithmic
$|\mathbf{p}|$-dependence of fermion velocity is actually robust in
the small-$|\mathbf{p}|$ region as the infrared cutoffs of
$\epsilon$ and $|\mathbf{p}|$ decrease. Different from the
small-$|\mathbf{p}|$ region, EPI can drive $v(\mathbf{p})$ to
deviate from the standard logarithmic behavior in
large-$|\mathbf{p}|$ region.

Once the full fermion propagator $G(p)$ is determined, one can
proceed to analyze the interaction effects on the properties of
bosons. According the analytical computations presented in Appendix
\ref{Sec:appb}, the DS equation of full phonon propagator $D(q)$ and
that of full $A$ boson propagator $F(q)$ are
\begin{eqnarray}
D(q) &=& D_{0}(q)\Big[1 - iN \int d p D_0(q) g^{2}\mathrm{Tr}
\big[G(p+q)\Gamma_0(q,p)G(p)\big]\Big], \label{eq:dsedq} \\
F(q) &=& F_{0}(q)\Big[1 - iN\int d p F_0(q)\mathrm{Tr} \big[G(p+q)
\Gamma_0(q,p)G(p)\big]\Big],\label{eq:dsefq}
\end{eqnarray}
which are derived from Eq.~(\ref{eq:dseofdq}) and
Eq.~(\ref{eq:dseoffq}), respectively. The equations of $D(q)$ and
$F(q)$ are no longer self-consistent, and can be directly computed
once the full fermion propagator $G(p)$ is obtained by solving its
DS equation. The polarization functions $\Pi_{p}(q)$ and
$\Pi_{A}(q)$, namely the self-energy functions of phonons and
Coulomb interaction, can also be calculated by using $D(q)$ and
$F(q)$, as shown by Eq.~(\ref{eq:polaphonon}) and
Eq.~(\ref{eq:polacoulomb}). More details about the interaction
effects on bosons can be found in Appendix \ref{Sec:appb}. While
these issues are interesting and deserve further investigations,
they apparently have no influence on the renormalization of fermion
velocity and will be addressed in separate works.

\section{Summary and Discussion \label{Sec:summary}}

In summary, here we present a non-perturbative study of the
interplay of EPI and Coulomb interaction in the context of graphene
by using the DS equation approach. In previous works, the effects of
EPI and Coulomb interaction are usually studied separately. When
both interactions are important, the situation becomes much more
involved. In this paper, we rigorously derive the DS equation of the
fully dressed Dirac fermion propagator $G(p)$ by taking into account
the interplay of EPI and Coulomb interaction. This equation is given
by Eq.~(\ref{eq:DSEGP}). As far as we know, such an equation has not
been obtained in previous publications. After carrying out a careful
analysis, we find that the correlation functions appearing in the DS
equation of $G(p)$ obey a number of exact identities, including
Eq.~(\ref{eq:d0gamma0}), Eq.~(\ref{eq:f0gamma0}), and
Eqs.~(\ref{eq:fourwtis}). All of these identities are derived from
the invariance of the partition function under various infinitesimal
changes of the fermionic and bosonic operators. Based on these
identities, we prove that the DS equation of $G(p)$ is indeed
self-closed. This is the main new result of this work.

As an application of our approach, we study how the fermion velocity
is renormalized. By numerically solving the self-closed DS equation
of $G(p)$ by means of iteration method, we show that the momentum
dependence and the energy dependence of the renormalized fermion
velocity is dominantly determined by the Coulomb interaction and the
EPI, respectively. In particular, the renormalized velocity
$v(\mathbf{p})$ exhibits a logarithmic $|\mathbf{p}|$-dependence
over a broad range of $|\mathbf{p}|$. This theoretical result is in
good agreement with the existing experiments of graphene
\cite{Elias11, Lanzara11, Chae12}.

We now comment on the range of applicability of our approach. To
make the DS equation of $G(p)$ self-closed, it is necessary to
derive a sufficient number of WTIs. In the model considered in this
work, there is only one coupling term for each FBI, namely
$\phi\bar{\psi}\gamma_{0}\psi$ for EPI and
$A\bar{\psi}\gamma_{0}\psi$ for Coulomb interaction. One can find
enough matrices to eliminate the special correlation functions
appearing in r.h.s. of Eq.~(\ref{eq:fivepoint}). For a FBI term that
has more than one components, it would be hard to eliminate such
correlation functions. Let us take relativistic QED$_{4}$
\cite{Itzykson} as an example. The Lagrangian density of QED$_{4}$
is given by
\begin{eqnarray}
\mathcal{L}_{\mathrm{QED}} = \sum_{\sigma}^{N}\bar{\psi}_{\sigma}
\gamma^{\mu}\left(i\partial_{\mu} - e a_{\mu}\right)\psi_{\sigma} -
\frac{1}{4}F^{\mu\nu}F_{\mu\nu},
\end{eqnarray}
where $\psi$ is a four-component spinor and $a_{\mu}$ is an abelian
gauge field. $F_{\mu\nu} = \partial_{\mu}a_{\nu} -
\partial_{\nu}a_{\mu}$ is the electricmagnetic tensor. Different
from EPI and Coulomb interaction, the gauge interaction term is
composed of four components, namely
$a_{\mu}\bar{\psi}\gamma^{\mu}\psi$ with $\mu=0,1,2,3$. Let the
spinor field transform as $\psi \to e^{i\theta \gamma^{m}}\psi$,
where $\theta$ is an infinitesimal constant and $\gamma^{m}$ could
be any $4\times 4$ matrix. On the basis of the invariance of the
partition function $Z$ under such transformations, one would obtain
an identity analogous to Eq.~(\ref{eq:variationaction}). Such an
identity would contain the following term
\begin{eqnarray}
\langle a_{\mu}(x){\bar\psi}_{\sigma}(x)(\gamma^{\mu}\gamma^{m} -
\gamma^{m}\gamma^{\mu})\psi_{\sigma}(x) \psi(y){\bar\psi}(z)\rangle.
\end{eqnarray}
There are not enough $\gamma^{m}$ matrices to fulfill the constraint
$\gamma^{\mu}\gamma^{m} - \gamma^{m}\gamma^{\mu}=0$ for all the four
components of $\gamma^{\mu}$. Thus the above correlation function
cannot be simply eliminated. It then becomes difficult to prove that
the DS equation of the full fermion propagator is self-closed. The
same difficulty also exists in QED$_{3}$. In fact, such a difficulty
is encountered in any quantum field theory in which the
fermion-boson coupling has two or more components. For instance,
when the spin degrees of freedom of Dirac fermions become important,
we need to consider such a coupling term as
$\psi^{\dag}\mathbf{\gamma}\cdot \psi \mathbf{S}$, where
$\mathbf{S}$ is a three-dimensional spin operator. We should further
generalize our approach to deal with these complicated models.

\section*{ACKNOWLEDGEMENTS}

We thank Jie Huang, Jing-Rong Wang, and Hao-Fu Zhu for helpful
discussions.

\appendix

\section{Derivation of the interaction vertex functions \label{Sec:appa}}

Here we show how to use the fermion and boson propagators (two-point
correlation functions) to express the following (connected)
three-point correlation function:
\begin{eqnarray}
\langle\phi(x)\psi_{\sigma}(y)\bar{\psi}_{\sigma}(z)\rangle_c &=&
\frac{\delta^3W}{\delta J(x)\delta\bar{\eta}_{\sigma}(y)
\delta\eta_{\sigma}(z)}\Big|_{J=0}.
\end{eqnarray}
According to the elementary rules of function integral
\cite{Itzykson}, we re-write the above expression as
\begin{eqnarray}
\frac{\delta^3W}{\delta J(x)\delta\bar{\eta}_{\sigma}(y)
\delta\eta_{\sigma}(z)}\Big|_{J=0} &=& \frac{\delta}{\delta
J(x)}\Big|_{J=0}\frac{\delta^2W}{\delta
\bar{\eta}_{\sigma}(y)\delta\eta_{\sigma}(z)}
\nonumber \\
&=& -\frac{\delta}{\delta J(x)}\Big|_{J=0}\left( \frac{\delta^2\Xi}
{\delta\bar{\psi}_{\sigma}(y)\delta\psi_{\sigma}(z)} \right)^{-1}
\nonumber \\
&=& \int dy^{\prime} dz^{\prime}\left(\frac{\delta^2\Xi}{\delta
\bar{\psi}_{\sigma}(y)\delta\psi_{\sigma}(y^{\prime})}\right)^{-1}
\nonumber \\
&& \times \left[\frac{\delta}{\delta J(x)}\Big|_{J=0}
\frac{\delta^2\Xi}{\delta\bar{\psi}_{\sigma}(y^{\prime})
\delta\psi_{\sigma}(z^{\prime})}\right]\left(\frac{\delta^2
\Xi}{\delta \bar{\psi}_{\sigma}(z^{\prime})\delta\psi_{\sigma}(z)}
\right)^{-1},\label{eq:deltajetaeta}
\end{eqnarray}
The operator $\frac{\delta}{\delta J(x)}\Big|_{J=0}$ appearing in
Eq.~(\ref{eq:deltajetaeta}) needs to be treated carefully. It can be
expanded as
\begin{eqnarray}
\frac{\delta}{\delta J(x)}\Big|_{J=0} &=& \frac{\delta\phi}{\delta
J}\Big|_{J=0}\frac{\delta}{\delta \phi}\Big|_{J=0} +\frac{\delta
A}{\delta J}\Big|_{J=0} \frac{\delta}{\delta A}\Big|_{J=0}
+\sum_{\sigma}^N\left(\frac{\delta \bar{\psi}_{\sigma}}{\delta
J}\Big|_{J=0}
\frac{\delta}{\delta\bar{\psi}_{\sigma}}\Big|_{J=0}+\frac{\delta
\psi_{\sigma}}{\delta
J}\Big|_{J=0}\frac{\delta}{\delta\psi_{\sigma}}\Big|_{J=0} \right)
\nonumber \\
&=& i\langle\phi\phi\rangle_c\frac{\delta}{\delta \phi}\Big|_{J=0}
+i\langle \phi A \rangle_c\frac{\delta}{\delta A}\Big|_{J=0} +
\sum_{\sigma}^N\left( \langle\phi\bar{\psi}_{\sigma}\rangle_c
\frac{\delta}{\delta \bar{\psi}_{\sigma}}\Big|_{J=0} +
\langle\phi\psi_{\sigma}\rangle_c
\frac{\delta}{\delta\psi_{\sigma}}\Big|_{J=0} \right) \nonumber \\
&=& -D\frac{\delta}{\delta \phi}\Big|_{J=0} -
D_F\frac{\delta}{\delta A}\Big|_{J=0}+0+0.
\end{eqnarray}
It is obviously true that $\langle\phi\bar{\psi}_{\sigma}\rangle_{c}
= \langle\phi\psi_{\sigma}\rangle_{c} = 0$ when all external sources
are removed because a fermion (boson) cannot be converted into a
boson (fermion) without inducing additional changes. However, one
cannot simply set $\langle \phi A \rangle_{c}=0$. Although there is
no direct coupling between $\phi$ and $A$ bosons in the Lagrangian
density (tree-level), they are both coupled to fermions and thus can
be turned into each other via quantum corrections (loop-level).
Phonons result from the lattice vibration and EPI basically
describes the mutual influence between negatively charged fermions
and positively charged ions. On the other hand, the Coulomb
interaction is experienced by negatively charged fermions. As the
ions are vibrating, the resultant phonon excitations affect the
surrounding electric field of fermions, which in turn alters the
Coulombic potential between fermions. These processes are embodied
in such correlation function as $\langle\phi A\rangle$ and $\langle
A \phi\rangle$. Substituting the above expression of
$\frac{\delta}{\delta J(x)}\Big|_{J=0}$ into
Eq.~(\ref{eq:deltajetaeta}) leads to
\begin{eqnarray}
&&\frac{\delta^3W}{\delta J(x)\delta\bar{\eta}_{\sigma}(y)
\delta\eta_{\sigma}(z)}\Big|_{J=0} \nonumber \\
&=& -\int dx^{\prime} dy^{\prime} dz^{\prime} \left(\frac{\delta^2
\Xi}{\delta\bar{\psi}_{\sigma}(y)\delta\psi_{\sigma}(y^{\prime})}
\right)^{-1} \nonumber \\
&&\times \left[\left(D(x-x^{\prime})\frac{\delta}{\delta
\phi(x^{\prime})}\Big|_{J=0}+D_F(x-x^{\prime})\frac{\delta}{\delta
A(x^{\prime})}\Big|_{J=0}\right)\frac{\delta^2\Xi}{\delta
\bar{\psi}_{\sigma}(y^{\prime})\delta\psi_{\sigma}(z^{\prime})}
\right]\left(\frac{\delta^2\Xi}{\delta\bar{\psi}_{\sigma}(z^{\prime})
\delta\psi_{\sigma}(z)} \right)^{-1} \nonumber \\
&=& -\int dx^{\prime}dy^{\prime}dz^{\prime}D(x-x^{\prime})
G(y-y^{\prime})\frac{\delta^3\Xi}{\delta\phi(x^{\prime})\delta
\bar{\psi}_{\sigma}(y^{\prime})\delta\psi_{\sigma}(z^{\prime})}
\Big|_{J=0}G(z^{\prime}-z) \nonumber \\
&& -\int dx^{\prime} dy^{\prime} dz^{\prime}D_F(x-x^{\prime})
G(y-y^{\prime}) \frac{\delta^3\Xi}{\delta A(x^{\prime})
\delta\bar{\psi}_{\sigma}(y^{\prime}) \delta\psi_{\sigma}
(z^{\prime})}\Big|_{J=0} G(z^{\prime}-z).\label{eq:deltaw}
\end{eqnarray}
The two-point correlation functions $G$, $D$, and $D_{F}$ are
already defined in Sec.~\ref{Sec:DSE}.

\section{Derivation of the DS equations of fermion and boson propagators \label{Sec:appb}}

We first derive the DS equation of the full fermion propagator
$G(p)$, taking into account the corrections from two different FBIs.
The partition function $Z$ is invariant under an arbitrary
infinitesimal change $\delta \bar{\psi}_{\sigma}$. This feature can
be used to obtain an identity
\begin{eqnarray}
\langle\left(i\partial_0\gamma_0 - i\partial_1 \gamma_1 - i
\partial_{2}\gamma_2\right)\psi_{\sigma}(x) -g\phi(x)\gamma_0\psi_{\sigma}(x)-
A(x)\gamma_0\psi_{\sigma}(x)+\eta_{\sigma}(x)\rangle = 0.
\end{eqnarray}
It is convenient to express this identity in terms of the generating
functional $W$ as follows
\begin{eqnarray}
&&(i\partial_0\gamma_0-i\partial_1\gamma_1-i\partial_2\gamma_2)
\frac{\delta W}{\delta\bar{\eta}_{\sigma}(x)} + ig\gamma_0
\frac{\delta^2W}{\delta J(x)\delta\bar{\eta}_{\sigma}(x)} -
g\gamma_{0} \frac{\delta W}{\delta J(x)}\frac{\delta W}{\delta
\bar{\eta}_{\sigma}(x)} \nonumber \\
&& +i\gamma_0\frac{\delta^2W}{\delta
A(x)\delta\bar{\eta}_{\sigma}(x)} -\gamma_0\frac{\delta W}{\delta
A(x)}\frac{\delta W}{\delta
\bar{\eta}_{\sigma}(x)}+\eta_{\sigma}(x)=0.
\end{eqnarray}
Applying the variation $-\frac{\delta}{\delta
\eta_{\sigma}(y)}\Big|_{J=0}$ to this identity, we obtain
\begin{eqnarray}
\delta(x-y) = (i\partial_0\gamma_0 - i\partial_1\gamma_1 -
i\partial_2\gamma_2) \frac{\delta^2W}{\delta
\bar{\eta}_{\sigma}(x)\delta\eta_{\sigma}(y)} + ig\gamma_0\langle
\phi(x)\psi_{\sigma}(x)\bar{\psi}_{\sigma}(y)\rangle_c +
i\gamma_0\langle A(x)\psi_{\sigma}(x)\bar{\psi}_{\sigma}(y)\rangle_c.
\nonumber \\
\end{eqnarray}
Substituting Eq.~(\ref{eq:expanded}) and Eq.~(\ref{eq:expanded2})
into it and making Fourier transformation give rise to
\begin{eqnarray}
1 &=& (p_0\gamma_0-\mathbf{\gamma\cdot p})G(p) -i\int_{q} g\gamma_0
D(q)G(p+q)\Gamma_p(q,p)G(p) -i\int_{q} g\gamma_0
D_F(q)G(p+q)\Gamma_A(q,p)G(p) \nonumber \\
&& -i\int_{q} \gamma_0 F_D(q)G(p+q)\Gamma_p(q,p)G(p) - i\int_{q}
g\gamma_0 F(q)G(p+q)\Gamma_A(q,p)G(p).
\end{eqnarray}
This equation can be re-written in a more compact form
\begin{eqnarray}
G^{-1}(p) &=& G_{0}^{-1}(p)-i\int_{q} g\gamma_0
G(p+q)D(q)\Gamma_p(q,p)-i\int_{q} \gamma_{0}G(p+q)F(q)\Gamma_A(q,p)
\nonumber \\
&& -i\int_{q} g \gamma_{0}G(p+q)D_F(q)\Gamma_A(q,p) - i\int_{q}
\gamma_{0}G(p+q)F_D(q)\Gamma_p(q,p). \label{eq:DSEGPapp}
\end{eqnarray}
Apparently, this equation is independent of the fermion flavor index
$\sigma$. In other words, the fermion propagator $G_{\sigma}(p)$ for
each flavor $\sigma$ satisfies the same DS equation.

Then we derive the DS equations satisfied by the full boson
propagators. As usual, we first work with the real time and the real
energy and finally replace the real energy with the imaginary
Matsubara frequency.

When one makes an arbitrary infinitesimal change of the phonon field
$\phi$, the partition function $Z$ should not change. This allows us
to obtain an equation
\begin{eqnarray}
\langle \mathbb{D}\phi(x) -ig \sum_{\sigma}^{N}
\bar{\psi}_{\sigma}(x)\gamma_{0}\psi_{\sigma}(x)\rangle + J(x) = 0,
\end{eqnarray}
which is equivalent to
\begin{eqnarray}
\mathbb{D}\frac{\delta W}{\delta J(x)} +i\sum_{\sigma}^{N}
g\mathrm{Tr}\left[\gamma_{0}\frac{\delta^{2} W}{\delta
\bar{\eta}_{\sigma}(x)\delta\eta_{\sigma}(x)}\right] + J(x)=0.
\end{eqnarray}
Perform a functional derivative $\frac{\delta}{\delta J(y)}|_{J=0}$
to both sides of this equation leads to
\begin{eqnarray}
&&\mathbb{D}D(x-y) + iN\int dz dz^{\prime} dy^{\prime}g
D(y-y^{\prime})\mathrm{Tr}[\gamma_0 G(x-z) \frac{\delta^{3}
\Xi}{\delta\phi(y^{\prime}) \delta\bar{\psi}_{\sigma}(z)\delta
\psi_{\sigma}(z^{\prime})}\Big|_{J=0} G(z^{\prime}-x) ] \nonumber\\
&& +iN\int dzdz^{\prime} dy^{\prime} gD_F(y-y^{\prime})
\mathrm{Tr}\left[\gamma_0 G(x-z) \frac{\delta^{3}
\Xi}{\delta\phi(y^{\prime}) \delta\bar{\psi}_{\sigma}(z)
\delta\psi_{\sigma}(z^{\prime})} \Big|_{J=0}
G(z^{\prime}-x)\right]-\delta(x-y) = 0.\nonumber \\
\end{eqnarray}
We now substitute Eq.~(\ref{eq:deltawdg}) into this equation and
then carry out Fourier transformations. The full phonon propagator
$D(q)$ is found to satisfy the following DS equation
\begin{eqnarray}
&& D_0^{-1}(q)D(q) + iN\int dp D(q)g\mathrm{Tr}[G(p+q)
\Gamma_p(q,p)G(p)] \nonumber \\
&& +iN\int dp D_F(q)g\mathrm{Tr}[G(p+q)\Gamma_A(q,p)G(p)] =1.
\end{eqnarray}
This DS equation is formally very complicated. Fortunately, using
the identity given by Eq.~(\ref{eq:d0gamma0}), we find that the DS
equation of $D(q)$ can be substantially simplified into
\begin{eqnarray}
D_{0}^{-1}(q)D(q) + iN \int d p D_0(q)
g^{2}\mathrm{Tr}[G(p+q)\Gamma_0(q,p)G(p)] = 1.
\label{eq:dseofdq}
\end{eqnarray}

\begin{figure}[htbp]
\centering
\includegraphics[width=5.1in]{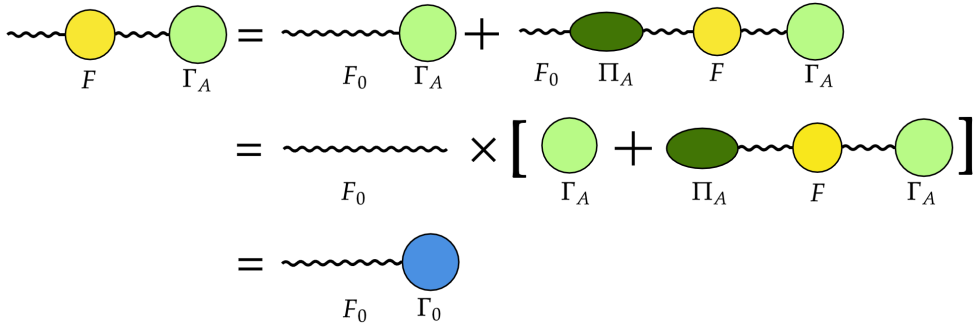}
\caption{A schematic illustration of the relation satisfied by
$\Gamma_{0}(q,p)$ and $\Pi_{A}(q)$.}\label{fig:pola}
\end{figure}

Then we apply the above derivational procedure to the other boson
field $A$ and, after repeating similar calculations, obtain the DS
equation of the full $A$ boson propagator $F(q)$:
\begin{eqnarray}
F_{0}^{-1}(q)F(q) + iN\int d p F_0(q)\mathrm{Tr}[G(p+q)
\Gamma_0(q,p)G(p)]=1. \label{eq:dseoffq}
\end{eqnarray}

The polarization functions, i.e., the self-energy functions, of the
phonons and the Coulomb interaction can be calculated as follows:
\begin{eqnarray}
\Pi_{p}(q) &=& D_{0}^{-1}(q) - D^{-1}(q),\label{eq:defipip}\\
\Pi_{A}(q) &=& F_{0}^{-1}(q) - F^{-1}(q).\label{eq:defipia}
\end{eqnarray}
Let us take $\Pi_{A}(q)$ as an example to illustrate how the current
vertex function $\Gamma_{0}(q,p)$ is related to the polarization
function. In the absence of phonons, the identity
Eq.~(\ref{eq:f0gamma0}) becomes
\begin{eqnarray}
F(q)\Gamma_A(q,p) = F_0(q)\Gamma_0(q,p).
\label{eq:fgammaaf0gamma0}
\end{eqnarray}
Making use of Eq.~(\ref{eq:defipia}), this identity is converted
into
\begin{eqnarray}
F(q)\Gamma_{A}(q,p) &=& \left[F_0(q) + F_0(q)\Pi_{A}(q)
F(q)\right]\Gamma_{A}(q,p) \nonumber \\
&=& F_{0}(q)\left[1+\Pi_{A}(q)F(q)\right]\Gamma_{A}(q,p) \nonumber \\
&=& F_{0}(q)\Gamma_0(q,p).\label{eq:ff0gammaagamma0}
\end{eqnarray}
This derivational process can be intuitively illustrated by the
diagrams plotted in Fig.~\ref{fig:pola}. It is now obvious that
$\Gamma_{0}(q,p)$ depends on $\Pi_{A}(q)$ via the relation
\begin{eqnarray}
\Gamma_0(q,p) = \left[1+\Pi_{A}(q)F(q)\right]\Gamma_{A}(q,p).
\end{eqnarray}
The fermion flavor $N$ enters into $\Pi_{A}(q)$ and also into
$F(q)$. However, $\Gamma_0(q,p)$ is independent of $N$, as shown by
Eq.~(\ref{eq:gamma0}). It can be inferred that the $N$-dependence of
$1+\Pi_{A}(q)F(q)$ cancels that of $\Gamma_{A}(q,p)$.

It is appropriate at this stage to transform real energy into
imaginary frequency. After doing so we re-write the two full boson
propagators as
\begin{eqnarray}
D(\omega_{n^{\prime}},\mathbf{q}) &=& D_{0}
(\omega_{n^{\prime}},\mathbf{q})\Big[1 + N\sum_{n}\int
\frac{d^2\mathbf{q}}{(2\pi)^2} D_0(\omega_{n^{\prime}},\mathbf{q})
g^2 \mathrm{Tr}\big[G(\epsilon_{n^{\prime}+n},\mathbf{p+q})
\Gamma_0(q,p)G(\epsilon_n,\mathbf{p})\big]\Big], \\
F(\omega_{n^{\prime}},\mathbf{q}) &=&
F_{0}(\omega_{n^{\prime}},\mathbf{q})\Big[1 + N\sum_{n}\int
\frac{d^2\mathbf{q}}{(2\pi)^2} F_0(\omega_{n^{\prime}},\mathbf{q})
\mathrm{Tr}\big[G(\epsilon_{n^{\prime}+n},\mathbf{p+q})\Gamma_0(q,p)
G(\epsilon_n,\mathbf{p})\big]\Big].
\end{eqnarray}
The full propagators can be used to compute the polarization
functions. Specifically, the polarization function for phonons is
\begin{eqnarray}
\Pi_{p}(\omega_{n^{\prime}},\mathbf{q}) &=& D_{0}^{-1}
(\omega_{n^{\prime}},\mathbf{q}) - D^{-1}(\omega_{n^{\prime}},
\mathbf{q}) \nonumber \\
&=& \frac{N\sum_{n}\int \frac{d^2\mathbf{q}}{(2\pi)^2}
g^{2}\mathrm{Tr}\big[G(\epsilon_{n^{\prime}+n},\mathbf{p+q})
\Gamma_0(q,p) G(\epsilon_n,\mathbf{p})\big]}{1+N\sum_{n}\int
\frac{d^2\mathbf{q}}{(2\pi)^2} D_0(\omega_{n^{\prime}},\mathbf{q})
g^2 \mathrm{Tr}\big[G(\epsilon_{n^{\prime}+n},\mathbf{p+q})
\Gamma_0(q,p) G(\epsilon_n,\mathbf{p})\big]},\label{eq:polaphonon}
\end{eqnarray}
and the polarization function for $A$ boson (Coulomb interaction) is
\begin{eqnarray}
\Pi_{A}(\omega_{n^{\prime}},\mathbf{q}) &=&
F_0^{-1}(\omega_{n^{\prime}},\mathbf{q}) -
F^{-1}(\omega_{n^{\prime}},\mathbf{q}) \nonumber \\
&=& \frac{N\sum_{n}\int \frac{d^2\mathbf{q}}{(2\pi)^2} \mathrm{Tr}
\big[G(\epsilon_{n^{\prime}+n},\mathbf{p+q}) \Gamma_0(q,p)
G(\epsilon_n,\mathbf{p})\big]}{1+N\sum_{n}\int
\frac{d^2\mathbf{q}}{(2\pi)^2} F_0(\omega_{n^{\prime}},\mathbf{q})
\mathrm{Tr}
\big[G(\epsilon_{n^{\prime}+n},\mathbf{p+q})\Gamma_0(q,p)
G(\epsilon_n,\mathbf{p})\big]}.\label{eq:polacoulomb}
\end{eqnarray}

Based on the above results, in principle it would be straightforward
to analyze the interaction effects on the behaviors of two bosons.
For instance, the dielectric constant $\varepsilon$ becomes a
function of energy and momentum, formally given by
\begin{eqnarray}
\varepsilon(\omega_{n^{\prime}},\mathbf{q}) &=& 1 -
F_0(\omega_{n^{\prime}},\mathbf{q})
\Pi_{A}(\omega_{n^{\prime}},\mathbf{q}) \nonumber \\
&=& \frac1{1+N\sum_{n}\int \frac{d^2\mathbf{q}}{(2\pi)^2}
F_0(\omega_{n^{\prime}},\mathbf{q})
\mathrm{Tr}\big[G(\epsilon_{n^{\prime}+n},\mathbf{p+q})
\Gamma_{0}(q,p)G(\epsilon_n,\mathbf{p})\big]}.
\end{eqnarray}
Moreover, one can investigate the properties of plasmon mode by
studying the polarization functions and compute the renormalized
phonon velocity $c_{s}$ based on the full phonon propagator $D(q)$.
From the technical perspective, it is difficult to perform such
calculations because one needs to first find an efficient numerical
method to translate the functions obtained using imaginary
frequencies into retarded and advanced functions that depend on real
energies.

\end{document}